\documentclass[pra,twocolumn,preprintnumbers,superscriptaddress,showpacs]{revtex4-1} 
\usepackage{graphicx}
\usepackage[small]{caption}
\usepackage{subcaption}
\usepackage{ragged2e}
\DeclareCaptionJustification{justified}{\justifying}
\usepackage{latexsym,epsfig,amssymb,amsfonts,amsmath,graphicx,bbm,bm}

\usepackage{amsmath,amsfonts,amsthm,amssymb}
\usepackage{setspace}
\usepackage{amsmath}
\usepackage{appendix}
\usepackage[ansinew]{inputenc}
\usepackage{bbm}
\usepackage{bm}
\usepackage{amsbsy}
\usepackage{dsfont} 
\usepackage{epsfig}
\usepackage{epstopdf}
\usepackage{dsfont}
\usepackage{enumerate}
\usepackage[resetlabels]{multibib}

\usepackage{color}
\definecolor{dred}{rgb}{.8,0.2,.2}
\definecolor{dyellow}{rgb}{.7,.7,.0}
\definecolor{ddred}{rgb}{.4,.0,.0}
\definecolor{dblue}{rgb}{.2,.2,.8}
 
\newcommand{\ii}{\mathrm{i}}

\newcommand{\dd}{\mathrm{d}}
\newcommand{\dds}{\hspace{-8pt}\mathrm{d}}

\newcommand{\ket}[1]{ |  #1 \rangle}

\newcommand{\ketss}[1]{ | \hspace{-2pt} #1 \rangle}
\newcommand{\bra}[1]{ \langle #1  |}
\newcommand{\brass}[1]{ \langle #1 \hspace{-2pt} |}

\newcommand{\outprod}[2]{\ket{#1}\bra{#2}}
\newcommand{\outprodss}[2]{\ketss{#1}\brass{#2}}

\newcommand{\tr}{\textrm{tr}}

\newcommand{\eqr}[1]{Eq.\ (\ref{#1})}
\newcommand{\fir}[1]{Fig.\ \ref{#1}}
\newcommand{\secr}[1]{Sec.\ \ref{#1}}

\newcommand{\an}[1]{\hat{#1}}
\newcommand{\cre}[1]{\hat{#1}^\dag}

\newcommand{\ee}{\textrm{e}}

\newcommand{\rr}{\mathbf{r}}

\newcommand{\PP}{\mathcal{P}}
\newcommand{\OO}{\mathcal{O}}

\newcommand{\EE}{\mathrm{E}}
\newcommand{\Var}{\mathrm{Var}}

\newcommand{\TT}{\mathcal{T}}
\newcommand{\ZZ}{\mathcal{Z}}

\begin{document}

\title{Thermometry of ultracold atoms via non-equilibrium work distributions}

\author{T.\ H.\ Johnson}
\email{tomihjohnson@gmail.com} 
\affiliation{Centre for Quantum Technologies, National University of Singapore, 3 Science Drive 2, 117543, Singapore}
\affiliation{Clarendon Laboratory, University of Oxford, Parks Road, Oxford OX1 3PU, United Kingdom}
\affiliation{Keble College, University of Oxford, Parks Road, Oxford OX1 3PG, United Kingdom}
\author{F.\ Cosco}
\affiliation{Turku Centre for Quantum Physics, Department of Physics and Astronomy, University of Turku, FI-20014 Turun yliopisto, Finland}
\affiliation{Clarendon Laboratory, University of Oxford, Parks Road, Oxford OX1 3PU, United Kingdom}
\affiliation{Dip.\ Fisica, Universit\`{a} della Calabria, 87036 Arcavacata di Rende (CS), Italy}
\author{M.\ T.\ Mitchison}
\email{marktmitchison@gmail.com}
\affiliation{Quantum Optics and Laser Science Group, Blackett Laboratory, Imperial College London, London SW7 2BW, United Kingdom}
\affiliation{Clarendon Laboratory, University of Oxford, Parks Road, Oxford OX1 3PU, United Kingdom}
\author{D.\ Jaksch}
\affiliation{Clarendon Laboratory, University of Oxford, Parks Road, Oxford OX1 3PU, United Kingdom}
\affiliation{Centre for Quantum Technologies, National University of Singapore, 3 Science Drive 2, 117543, Singapore}
\affiliation{Keble College, University of Oxford, Parks Road, Oxford OX1 3PG, United Kingdom}
\author{S.\ R.\ Clark}
\affiliation{Clarendon Laboratory, University of Oxford, Parks Road, Oxford OX1 3PU, United Kingdom}
\affiliation{Keble College, University of Oxford, Parks Road, Oxford OX1 3PG, United Kingdom}
\affiliation{Department of Physics, University of Bath, Claverton Down, Bath BA2 7AY, United Kingdom}
\affiliation{Max Planck Institute for the Structure and Dynamics of Matter, Luruper Chaussee 149, 22761 Hamburg, Germany}

\date{\today}

\begin{abstract}
Estimating the temperature of a cold quantum system is difficult. 
Usually, one measures a well-understood thermal state and uses that prior knowledge to infer its temperature.
In contrast, we introduce a method of thermometry that assumes minimal knowledge of the state of a system and is potentially non-destructive.
Our method uses a universal temperature-dependence of the quench dynamics of an initially thermal system coupled to a qubit probe that follows from the Tasaki-Crooks theorem for non-equilibrium work distributions. 
We provide examples for a cold-atom system, in which our thermometry protocol may retain accuracy and precision at subnanokelvin temperatures.
\end{abstract}

\pacs{
06.20.-f, 
05.70.Ln, 
03.65.Yz, 
67.85.Pq} 

\maketitle 

\section{Introduction}

Many technological applications utilizing quantum systems, e.g.\ analogue quantum simulators~\cite{Johnson2014}, require precise and accurate measurements of their temperature, making thermometry of quantum systems a fundamental task. 
Conventional thermometry proceeds by bringing a small probe of known temperature-dependence into equilibrium with a thermal system and then measuring that probe. For accurate thermometry, this requires that the characteristic energy of the probe is precisely known and tuned near the thermal energy~\cite{Correa2015}. This can be challenging at the low temperatures relevant for experiments on ultracold atoms.

Instead, the temperature of ultracold gases is often inferred by directly measuring observables whose temperature-dependence is well understood. 
For instance, time-of-flight imaging of the momentum distribution is used to obtain the temperature of weakly interacting cold  atoms~\cite{Leanhardt2003,Bloch2008}. Meanwhile, for strongly interacting atoms in a lattice, temperature is inferred from fluctuations of lattice-site occupation numbers~\cite{Sherson2010}. However, such an approach is only feasible if the system Hamiltonian or its thermal states are well characterized and sufficiently simple, so that the temperature-dependence of observables can be calculated and compared with measurements. Unfortunately, these requirements are frequently unmet in cold-atom experiments, where the system can be strongly correlated and established perturbative or numerical techniques typically fail. What is missing is a {\em generic} approach for thermometry of cold atoms that does not need prior understanding of the thermal state.

To fill this gap we return to the idea of bringing probes into contact with an initially thermal quantum system, this time focusing on the ensuing {\em non-equilibrium} dynamics. This increasingly studied~\cite{Goold2011,Haikka2012,Sindona2013,Punk2013} and potentially non-destructive approach to investigating quantum systems has been used to analyze the parameters~\cite{Recati2005} and spectrum~\cite{Johnson2011} of a Hamiltonian, and the non-Markovianity of an open quantum system~\cite{Haikka2013}. Applications of this approach to the thermometry of cold atoms~\cite{Bruderer2006,Sabin2014,Hangleiter2015,Fedichev2003} show that focusing on non-equilibrium dynamics can avoid the requirement of precisely tuning the characteristic energy of the probe near the thermal energy. However, the particular approaches put forward so far rely on the system having a well-understood Hamiltonian.

In this article, we show how to use a non-equilibrium probe to infer the temperature of a cold-atom system, which may in principle have an \textit{arbitrary} Hamiltonian. Our approach exploits the Tasaki-Crooks theorem~\cite{Tasaki2000,Talkner2007,Campisi2011,Goold2015}: a universal temperature-dependent relationship between non-equilibrium work distributions that may be embedded in the state of a qubit probe~\cite{Dorner2013,Mazzola2013,Batalhao2014}. We demonstrate that this versatile method is naturally suited to thermometry of cold atomic gases, and is both accurate and robust in the presence of imperfect data. Importantly, no detailed knowledge of the internal dynamics of the system is needed. The only requirement is control over the coupling between the system and the two states of the qubit thermometer, which is readily achieved by using an atomic impurity as the probe. Our protocol thus realizes near-ideal thermometry within its domain of applicability, which corresponds to temperatures commensurate with or lower than the characteristic energy scales of the system. This is precisely the temperature regime of greatest interest for cold-atom physics, and also a challenging one for conventional thermometry.


We gauge the accuracy and precision of our protocol by simulating its application to a paradigmatic ultracold-atom system. Specifically, we consider a Bose-Hubbard model (BHM) and localized impurity qubit, as could be realized by cold bosons in an optical lattice and, for example, two internal states of a separately-trapped atom of a different species~\cite{Palzer2009,Will2011,Spethmann2012,Catani2012}. Our protocol maintains accuracy and precision to a few percent in all regimes investigated. This includes when the thermal energy is one or two orders of magnitudes lower than the hopping energy of the BHM, where the latter might typically correspond to tens of nanokelvin. Moreover, it includes intermediate interaction strengths for which the nature of the thermal state is poorly understood and neither time-of-flight nor number-fluctuation measurements reveal the temperature. This work thus opens the door for thermometry of generic cold-atom systems at extreme temperatures and the technologies, e.g.\ quantum simulation, that require such thermometry. We begin with a general description of our scheme, before proceeding to a detailed study of its application to the BHM.

\section{Description of the protocol}

\subsection{Non-equilibrium work distributions}
 Our thermometry protocol is based on a relationship between distributions of the work done by quenching a system away from equilibrium. We write $P_Q(W)$ for the distribution of the work $W$ done on a system, e.g.\ a cold atomic gas, due to a quench $Q$. In the quench, the parameter $\lambda$ appearing in a system's Hamiltonian $\hat{H} (\lambda) = \hat{H}_S +  \lambda \hat{V}$ is varied as $\lambda_Q (t)$ for $t \in [0, \tau]$ driving the system away from an initial thermal state $\hat{\rho}_\beta (\lambda_Q(0))$. Here $\beta$ is the inverse temperature, $\hat{\rho}_\beta (\lambda) = \ee^{-\beta \hat{H} (\lambda)} / \ZZ_\beta (\lambda)$ is a thermal state of the system and $\ZZ_\beta (\lambda) = \tr \{ \exp [ - \beta \hat{H} (\lambda) ]\}$ is the corresponding partition function.

The forward distribution $P_F(W)$ for some quench $\lambda_F(t)$ from $\lambda_i$ to $\lambda_f$ is related to the backward distribution $P_B(W)$ of its reverse $\lambda_B(t) = \lambda_F(\tau-t)$ from $\lambda_f$ to $\lambda_i$ via the Tasaki-Crooks relation~\cite{Tasaki2000,Talkner2007,Campisi2011,Goold2015}
\begin{equation}
\label{eq:CrooksTasaki}
\ln \{ R(W) \} = \ln \left \{ \frac{P_F(W)}{P_B(-W)} \right \} = \beta(W-\Delta F) .
\end{equation}
The ratio $R(W)$ of the work distributions therefore depends only on $\beta$ and one other constant, the free energy difference $\Delta F = F(\lambda_f) - F(\lambda_i)$, with $F(\lambda) = \beta^{-1} \ln \left [ \mathcal{Z}_\beta(\lambda)\right ]$. Note that this relation holds generally for a coherent quench; it is not based on assumptions of linear response or adiabaticity.

\subsection{Qubit interferometry}

To directly measure quantum work distributions~\cite{Sindona2014,Roncaglia2014,DeChiara2015}, and thus their ratio $R(W)$, requires overcoming significant challenges, namely the realization of often prohibitively large numbers of difficult projective energy measurements~\cite{Huber2008,An2014}. We instead consider an indirect approach to measuring $R(W)$ using qubit interferometry~\cite{Dorner2013,Mazzola2013,Batalhao2014}. The system of interest is brought into contact with a probe qubit, giving total Hamiltonian $\hat{H}_T(t) = -(\Delta/2) \hat{\sigma}_z + \hat{H}_S + \hat{H}_I(t)$. Here $\Delta$ is the difference in energy between the ground and excited states $\ketss{\downarrow}$ and $\ketss{\uparrow}$ of the qubit, $\hat{H}_S$ is the Hamiltonian of the system of interest, and the interaction $\hat{H}_I (t)$ takes the form
\begin{equation}
\label{eq:Interaction}
\hat{H}_I (t) = \left( g_\downarrow (t) \outprodss{\downarrow}{\downarrow} + g_\uparrow (t) \outprodss{\uparrow}{\uparrow} \right ) \otimes \hat{V} .
\end{equation}
The combined system is initialized at time $t=0$ in the state $\hat{\rho} = \outprod{s}{s} \otimes \hat{\rho}_\beta (\lambda_Q(0))$, with the qubit in some superposition $\ket{s} = s_\downarrow \ketss{\downarrow} + s_\uparrow \ketss{\uparrow}$. This could be achieved by first reaching equilibrium with $\beta \Delta \gg 1$ and $g_\downarrow (0) = g_\uparrow (0) = \lambda_Q(0)$, then applying a rotation $\hat{\sigma}_s = s_\uparrow \hat{\sigma}_x +  s_\downarrow \hat{\sigma}_z$ to the qubit. The state-dependent couplings $g_\downarrow (t)$ and $g_\uparrow (t)$ are then both varied according to the quench $\lambda_Q(t)$ to be investigated, but with the latter delayed by a time $u$, i.e.\
\begin{align}
\label{eq:CouplingsProtocol}
g_\downarrow(t) & = \left\lbrace \begin{array}{ll} \lambda_Q(t), \quad \qquad \! &  0 \leq t \leq \tau, \\ \lambda_Q (\tau), & \tau \leq t\leq \tau + u ,
\end{array}\right. \nonumber \\
g_\uparrow(t) & =  \left\lbrace \begin{array}{ll} \lambda_Q (0),  &  0 \leq t \leq u, \\ \lambda_Q(t - u), \quad & u \leq t \leq \tau + u.
\end{array}\right. \nonumber
\end{align}
At time $\tau + u$, when both quenches are complete, the qubit has a reduced density operator, in units where $\hbar = 1$,
\begin{equation}
\label{eq:FinalState}
\begin{aligned}
\hat{\rho}_q =& |s_\downarrow|^2 \outprodss{\downarrow}{\downarrow} + s^\ast_\uparrow s_\downarrow \ee^{\ii \Delta (\tau + u)} \chi_Q^\ast(u) \outprodss{\downarrow}{\uparrow} \\
&+ s^\ast_\downarrow s_\uparrow \ee^{-\ii \Delta (\tau + u)} \chi_Q(u) \outprodss{\uparrow}{\downarrow} + |s_\uparrow|^2 \outprodss{\uparrow}{\uparrow} .
\end{aligned}
\end{equation}
Here we have introduced the dephasing function
\begin{equation}
\label{eq:CharacteristicFunction}
\chi_Q (u) = \tr \left\{ \hat{U}^\dagger_Q \ee^{\ii u \hat{H}(\lambda_Q(\tau))} \hat{U}_Q \ee^{-\ii u \hat{H}(\lambda_Q(0))} \hat{\rho}_\beta (\lambda_Q(0)) \right\} , \nonumber
\end{equation}
where $\hat{U}_Q = \TT \exp [ - \ii \int_0^{\tau} \dd t \hat{H} (\lambda_Q(t)) ]$ evolves the system according to the time-dependent quench Hamiltonian $\hat{H} (\lambda_Q(t))$ and $\TT$ is the time-ordering operator.

Close examination reveals that $\chi_Q (u)$ is none other than the characteristic function, or Fourier transform, of the work distribution~\cite{Dorner2013,Mazzola2013}
\begin{equation}
P_Q (W) = (2 \pi)^{-1} \int \dd u \ee^{- \ii W u} \chi_Q (u) . \nonumber
\end{equation}
Hence it is possible to measure $P_Q (W)$ from $\chi_Q (u)$, and the latter from expected values
\begin{equation}
\label{eq:ExpectedValue}
\langle \hat{\sigma}_x \rangle + \ii \langle \hat{\sigma}_y \rangle = \tr_q \{ ( \hat{\sigma}_x + \ii \hat{\sigma}_y) \hat{\rho}_q \} = 2 s^\ast_\downarrow s_\uparrow \ee^{-\ii \Delta (\tau + u)} \chi_Q (u) , \nonumber
\end{equation}
for the qubit state [\eqr{eq:FinalState}] at the end of the interference protocol. In what follows, we set $s^\ast_\downarrow s_\uparrow = 1/2$ and $\Delta = 0$, with the more general case treated in the Supplemental Material.

\subsection{Thermometry} The main result of this article is that we are able to use the above relations to systematically, precisely and accurately infer temperature from realistically noisy data, without any knowledge regarding the thermal states $\hat{\rho}_\beta (\lambda)$ of the quantum system. One needs only to have good control over a single qubit and its interaction with the system, which can be achieved by using an atomic impurity as the probe.

Let us analyze these claims in order. Our protocol is robust in the presence of two fundamental sources of error, analyzed in detail in the Supplemental Material. First, it is only possible to estimate $\chi_Q (u)$ for a finite number of times $u_j$. Here we assume $N_{\mathrm{steps}}$ times $u_j = j T / N_{\mathrm{steps}}$ for $j = 1,...,N_{\mathrm{steps}}$. Second, each estimate of $\langle \hat{\sigma}_x \rangle$ and $\langle \hat{\sigma}_y \rangle$ used to estimate $\chi_Q (u_j)$ will have some error. Here we assume that errors are due to the finite number $N_{\mathrm{meas}}$ of measurements used to estimate each expectation value. However, other known qubit measurement errors can be treated in the same framework.

The first point means that rather than estimating $P_Q (W)$, we instead estimate $p_Q (W)$, a copy that is subject to spectral leakage, due to the finite time-window $T$, and aliasing, due to the discreteness of $u_j$. 
We show that provided typically modest values of $T$ and $N_{\mathrm{steps}}$ are chosen such that $\pi \beta / T$ is on the order of unity and $T / N_{\mathrm{steps}} \ll \tau_{Q\mathrm{deph}} = 1/\sigma_Q$, then any effect on the ratio is negligible i.e.\ $R(W) \approx p_F (W) / p_B (-W)$ (see Supplemental Material). Here $\tau_{Q\mathrm{deph}}$ is the delay $u$ needed for the qubit to significantly dephase and is the inverse of $\sigma_Q$, the width of the work distribution $P_Q(W)$. 

The second point means our estimate of $\chi_Q (u_j)$ will be an unbiased Gaussian random variable with variance $(2 - |\chi_Q (u_j)|^2)/N_{\mathrm{meas}}$, which then propagates linearly into an unbiased Gaussian estimate of $p_Q(W)$ with variance scaling as $T^2 / N_{\mathrm{steps}} N_{\mathrm{meas}}$. 
In a Bayesian approach detailed in the Supplemental Material, we show how this knowledge, together with the Tasaki-Crooks relation and the non-negativity of the work distributions, can be used to build the probability distribution $\PP(\beta)$ for $\beta$ given a set of estimates of $\chi_Q (u_j)$. It is also possible to include any prior knowledge of $\beta$ and $\Delta F$, though here we assume no prior knowledge. Our approach is found to be well calibrated and accurate, to a few percent, given a modest number of times $N_{\mathrm{steps}}$ and measurements $N_{\mathrm{meas}}$.

The universality of the Tasaki-Crooks temperature-dependence allows the thermometry protocol to be applied in complete ignorance of the quantum system's thermal state $\hat{\rho}_\beta (\lambda)$ and even its Hamiltonian $\hat{H}_S$. It only needs to be ensured that both states of the impurity couple to the same operator $\an{V}$ [\eqr{eq:Interaction}], that the coupling strengths $g_{\downarrow}(t)$ and $g_{\uparrow}(t)$ trace the same path with some delay, and that the backwards path mirrors the forward path. Such properties may be understood theoretically in advance or confirmed experimentally by observing how the qubit, in eigenstate $\ketss{\downarrow}$ or $\ketss{\uparrow}$, behaves when interacting with the system. The choices of perturbing Hamiltonian $\an{V}$ and quench $\lambda_Q (t)$ used for thermometry are arbitrary in principle, since they affect $\Delta F$ not $\beta$. In fact, provided the relationships above hold, the actual values of $\an{V}$ and $\lambda_Q (t)$ need not be known.

In practice, any thermometer benefits from optimization, and our protocol is no exception. In particular, the quench should be chosen so that the fluctuations of the non-equilibrium work are on the order of the temperature or larger. This ensures that the ratio of work distributions, $R(W)$ in Eq.~\eqref{eq:CrooksTasaki}, can be accurately and precisely inferred over a large enough range of values of the work $W$ to extract a good straight-line fit for $\beta$. Faced with an unknown system, the experimenter may therefore need to make some adjustments in order to find an appropriate quench. We emphasise that this optimization can be based purely on observations of the qubit evolution, without resorting to direct measurements on the system of interest. Nevertheless, it is clearly important to use a probe qubit whose interaction with the system is highly controllable and well understood over a range of energies.

In the context of cold atomic gases, a probe qubit comprising an atomic impurity satisfies these requirements well. In the following, we consider density-density coupling, as appropriate for an impurity interacting with a host gas of a different species. At low temperatures, this interaction is characterized by a small number of parameters, such as $s$-wave scattering lengths, which may be accurately measured via independent experiments and controlled by means of external fields. Furthermore, the range of interaction energies between the impurity atom and the atomic gas naturally coincides with the characteristic energies of interaction between the gas atoms themselves. It is therefore apparent that the non-equilibrium work fluctuations induced by an atomic impurity are well suited to characterize temperatures that are on the same order or smaller than the natural energy scales of the cold-atom system.


\section{Example with cold atoms in a lattice} For concreteness, from here on we take our system to be a cold atomic gas confined by an optical lattice, and the qubit to be formed by two internal states of an impurity atom of a different species, which a strong trap localizes to density $n_q(\rr)$~\cite{Palzer2009,Will2011,Spethmann2012,Catani2012}. Both impurity states $\ketss{\downarrow}$ and $\ketss{\uparrow}$ couple via the contact interaction to the weighted density operator $\hat{V} = \int \dd \rr n_q (\rr) \hat{\Psi}^\dagger (\rr) \hat{\Psi} (\rr)$ with different interaction strengths $g_{\downarrow}$ and $g_{\uparrow}$. Here $\hat{\Psi}^\dagger (\rr)$ and $\hat{\Psi} (\rr)$ are the field operators for the atoms comprising the system. Thus it is possible to realize a combined Hamiltonian $\hat{H}_T$ of the form required for our protocol. The qubit gate $\hat{\sigma}_s$, and the measurement of $\hat{\sigma}_x$ and $\hat{\sigma}_y$ can be performed e.g.\ using Rabi pulses combined with state-dependent fluorescing.

The separate time-dependence of the couplings strengths $g_{\downarrow}(t)$ and $g_{\uparrow}(t)$ could be controlled by Feschbach resonances~\cite{Palzer2009,Will2011,Spethmann2012,Catani2012}. Another option is to realize the time-dependence of the coupling strengths effectively by changing the properties of their trap, which may be state-selective. A further possibility is to forgo using internal states to form the qubit, instead splitting the wavefunction of an impurity, passing two copies through the system, and then interfering them~\cite{Schaff2014}.

For our specific example, we consider a one-dimensional Bose gas in a simple periodic lattice, which reduces to the Bose-Hubbard model~\cite{Jaksch1998}
\begin{align}
\label{BoseHubbardHamiltonian}
\an{H}_S = -J  \sum_{\langle j j' \rangle} \cre{a}_j \an{a}_{j'} + \sum_{j= 1}^M \left( \frac{U}{2} \cre{a}_j \cre{a}_j \an{a}_j \an{a}_j - \mu \cre{a}_j \an{a}_j \right) . 
\end{align}
Here $\cre{a}_j$ and $\an{a}_j$ create and annihilate a particle at site $j$ of a total of $M$, the hopping and interaction energies and chemical potential are written $J$, $U$ and $\mu$, respectively, and $\langle j j' \rangle$ represents a sum over nearest neighbors. Assuming the impurity to be localized at a single central site $c$, the system's interaction Hamiltonian is $\an{V} = \eta \cre{a}_c \an{a}_c$, where $\eta = \int \dd \rr n_q (\rr) |w_c (\rr)|^2$ and $w_c(\rr)$ is the Wannier function at the central site.


\begin{figure}[tb]
        \begin{subfigure}[b]{1.68 in}
                \includegraphics[width=\textwidth]{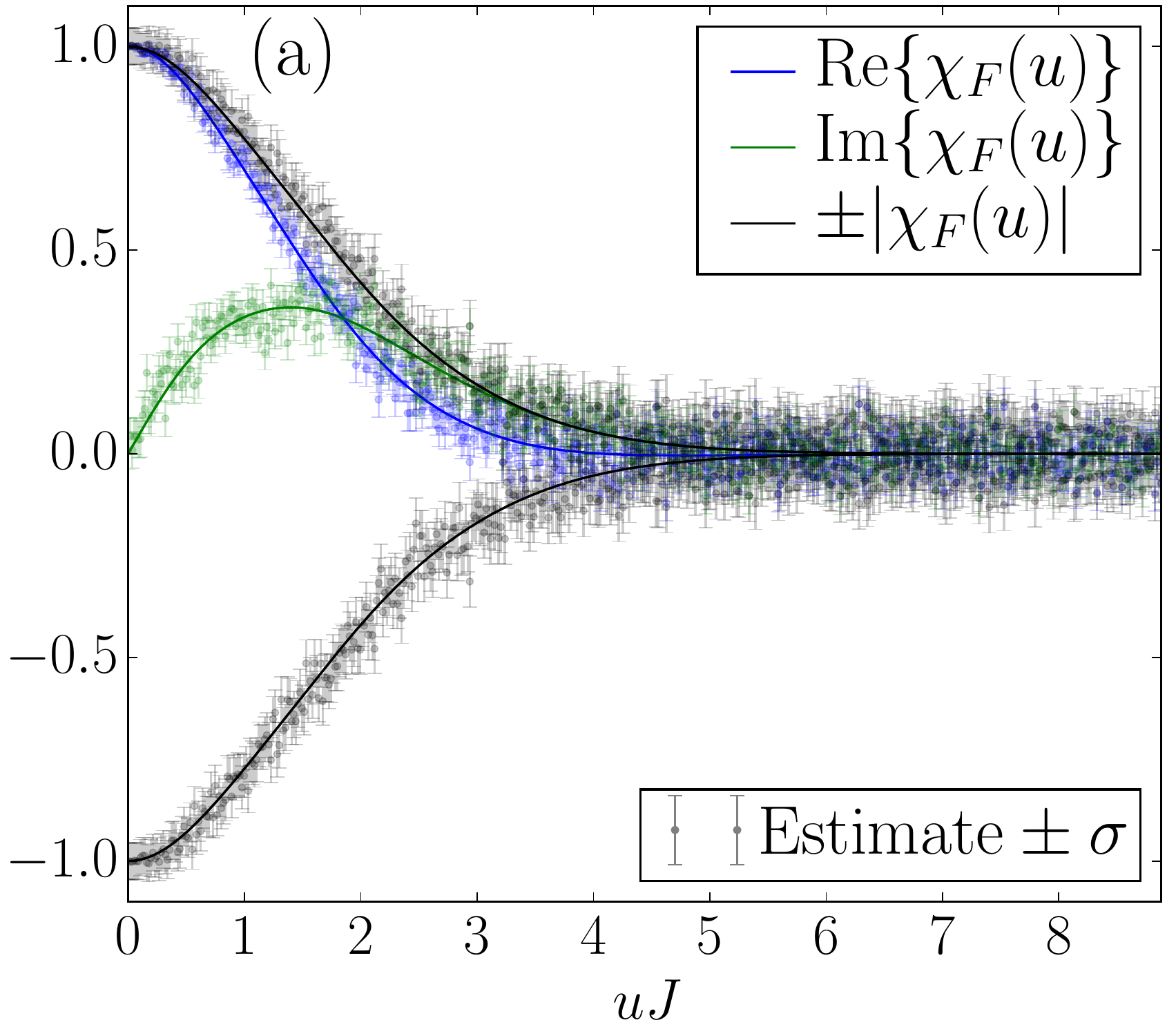}
                \label{fig:chiFvsu}
                \vspace{-10pt}
        \end{subfigure}
        \begin{subfigure}[b]{1.68 in}
                \includegraphics[width=\textwidth]{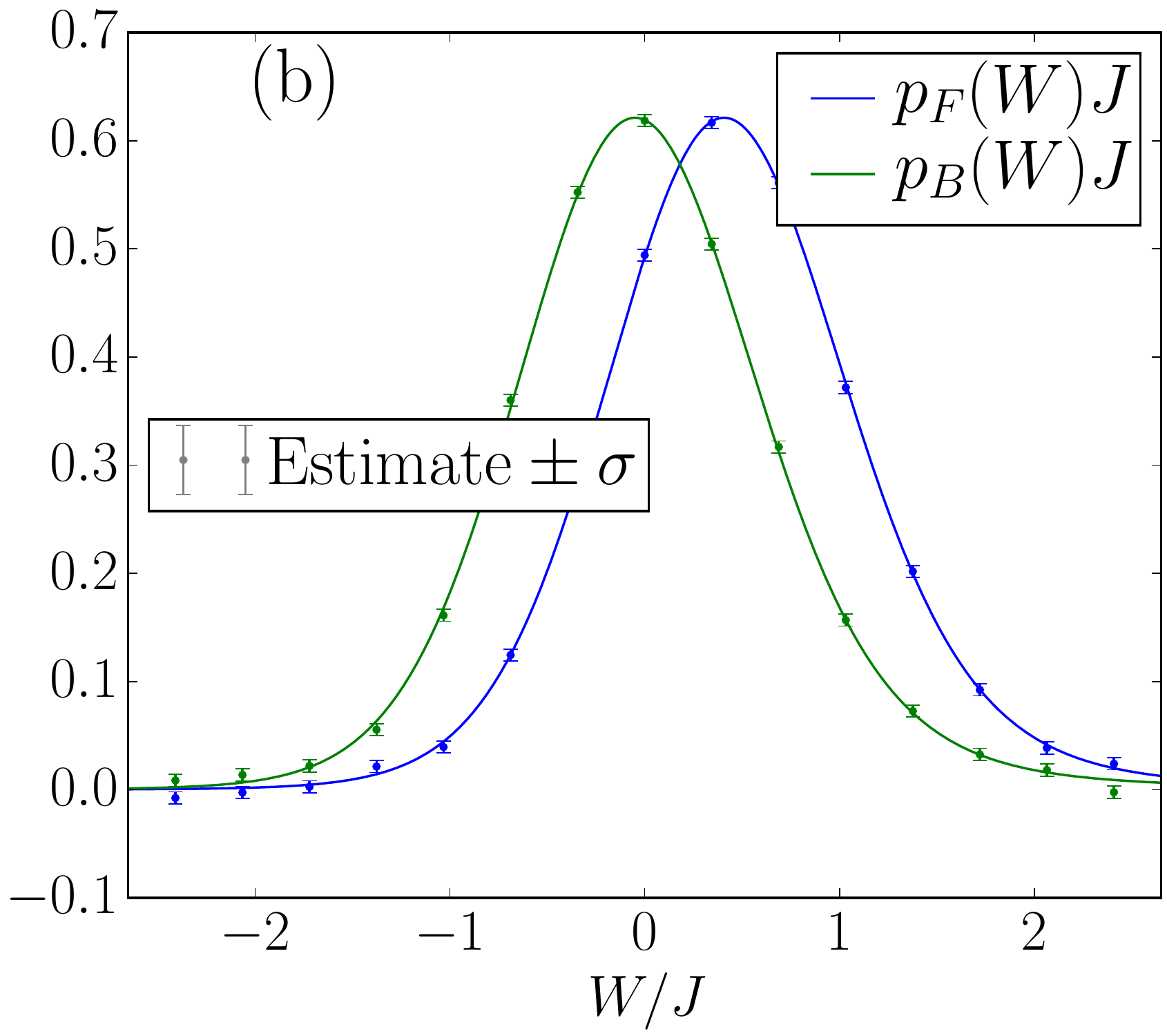}
                \label{fig:PsvsW}
                \vspace{-10pt}
        \end{subfigure}
        \begin{subfigure}[b]{1.68 in}
                \includegraphics[width=\textwidth]{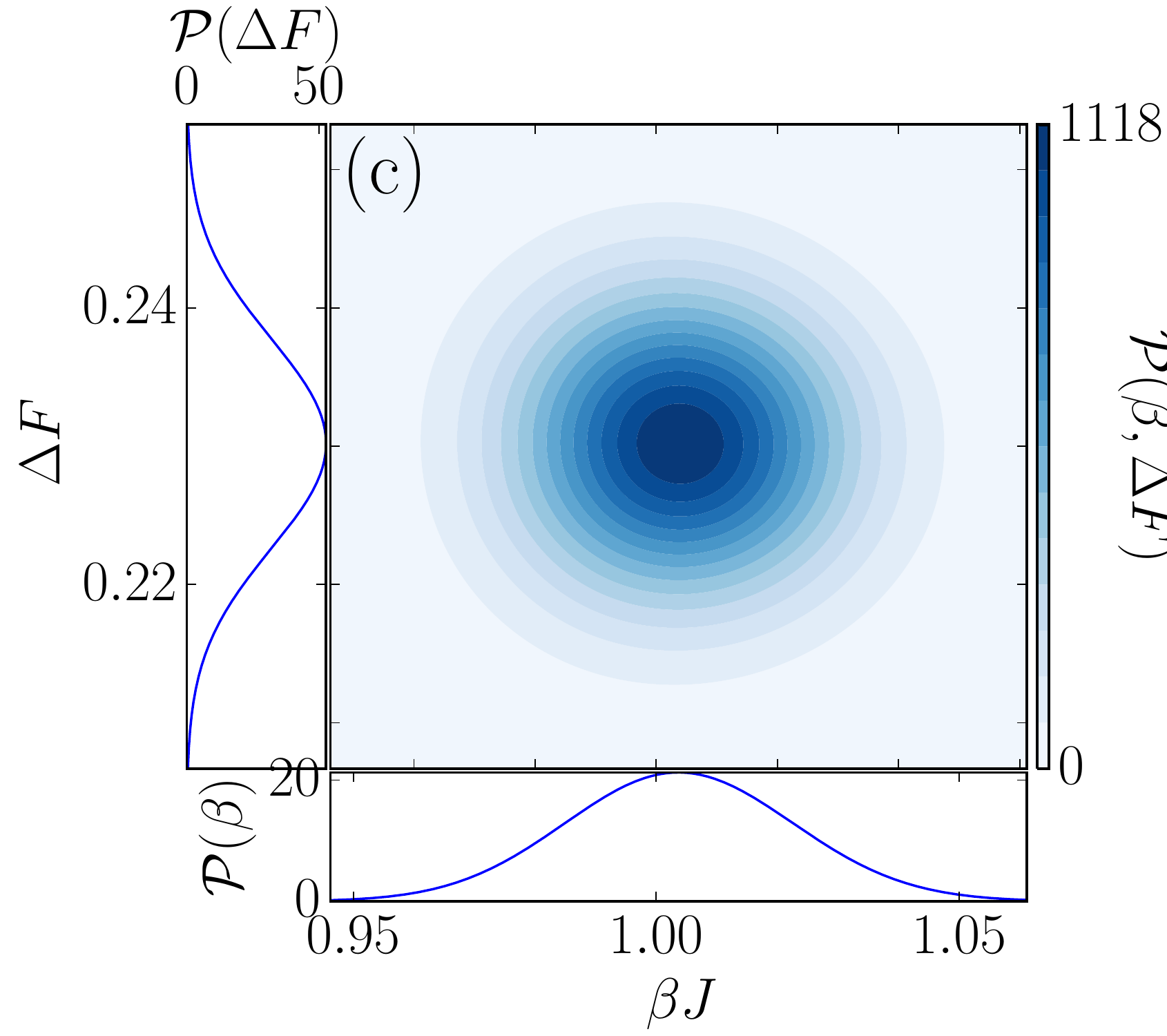}
                \label{fig:PsRatio}
        \end{subfigure}
        \begin{subfigure}[b]{1.68 in}
                \includegraphics[width=\textwidth]{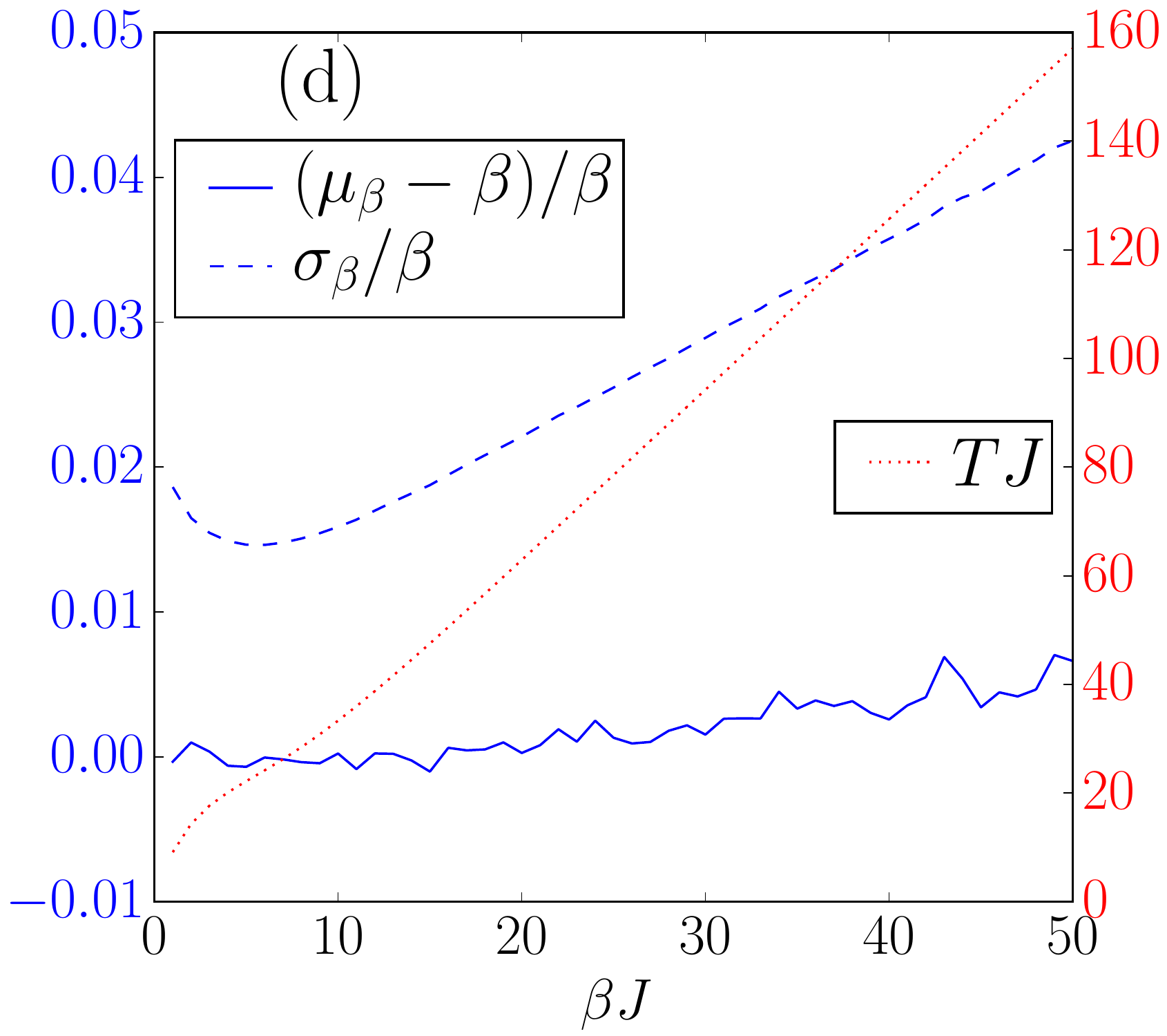}
                \label{fig:lowtemp}
        \end{subfigure}
        \caption{\label{fig:typicaltemp} {\em Superfluid phase}. (a) Characteristic function $\chi_F(u)$ of the forward quench. (b) Work distributions $p_Q(W)$. (c) The joint distribution $\PP (\beta, \Delta F)$ inferred from the estimates of $p_Q(W)$, together with the corresponding marginal distributions $\PP (\beta)$ and $\PP (\Delta F)$. 
(d) The fractional standard deviation $\sigma_\beta /\beta$ and bias $(\mu_\beta - \beta) / \beta$ of the $\beta$ estimates observed over 1000 simulated experiments, plotted against the actual $\beta$. The values of $T$ are shown in the figure. Unless stated otherwise, for all figures, the system parameters are $M = 1000$, $U/J = 0.1$ and $n = 1$, the quench parameters are $\lambda_i \eta / J = 0$, $\lambda_f \eta /J = 0.5$, and $\tau J =1$, and the protocol parameters are $N_{\mathrm{steps}} = 500$, $N_{\mathrm{meas}} = 500$, and $T J = 2.9 \pi$.}
\end{figure}


\subsection{Superfluid phase}

In the superfluid regime $nU/J \ll 1$, with $n$ the number of bosons per site, we can describe the system approximately in terms of phononic Bogoliubov excitations above the uniform condensate of density $n$. To second order in these excitations and terms that create them, the system and interaction Hamiltonians simplify~\cite{VanOosten2001}, up to a constant, to
\begin{align}
\an{H}_S = & \sum_k \omega_k \cre{b}_k \an{b}_k ,  \nonumber \\
\an{V} = & \eta n + \sum_k ( \eta^\ast_k \cre{b}_k + \eta_k \an{b}_k ) .  \nonumber
\end{align}
Here, $\cre{b}_k$ and $\an{b}_k$ create and annihilate a phonon at quasimomentum $k$, $ \omega_k = \sqrt{ \epsilon_k ( \epsilon_k + 2 U n )}$ is the phonon dispersion written in terms of single-particle energies $\epsilon_k = 2J(1 - \cos ka)$, and $\eta_k = \eta \sqrt{n \epsilon_k / M \omega_k} \ee^{-\ii k a c}$ is the relative coupling of each phonon mode to the impurity, with $a$ the lattice parameter.

With the Hamiltonians $\an{H}_S$ and $\an{V}$ in this form, the characteristic function $\chi_Q (u)$ can be calculated exactly (see Supplemental Material) from time-integrals of the form $\Lambda_{Q k} = \omega_k \int_0^{\tau} \dd t \lambda_Q (t) \exp^{-\ii \omega_k t}$. The specific quench considered here is $\lambda_Q (t) = \lambda_Q (0) + (\lambda_Q (\tau) - \lambda_Q (0)) \sin^2 (\pi t / 2 \tau)$.

We demonstrate the protocol first for a temperature corresponding to the typical energy scale of the system, $\beta J = 1$, with the results shown in \fir{fig:typicaltemp}. In \fir{fig:typicaltemp}(a) we have plotted the known characteristic function values $\chi_F (u_j)$ for the forward quench. Also shown are the estimated values and associated errors from a single simulated experiment, consisting of $2 N_{\mathrm{meas}}$ measurements at $N_{\mathrm{steps}}$ times. In \fir{fig:typicaltemp}(b) we show the corresponding work distributions, both forward $p_F(W)$ and backward $p_B(W)$, obtained from exact and estimated values of $\chi_F (u_j)$ and $\chi_B (u_j)$, again with error bars.
Figure \ref{fig:typicaltemp}(c) shows the joint distribution $\PP (\beta, \Delta F)$ of $\beta$ and $\Delta F$ conditioned upon the estimates of $p_Q(W)$ obtained. From this, the marginal distribution $\PP (\beta)$ for $\beta$, also shown, is calculated. The distribution of this example is consistent with the known value, containing uncertainty in $\beta$ of only a few percent. This can be reduced by increasing $N_{\mathrm{steps}}$ or $N_{\mathrm{meas}}$.

The Bayesian prediction is remarkably well calibrated. For each of several values of $\beta$ between $1$ and $50$, we have simulated $1000$ experiments of the above type. In \fir{fig:typicaltemp}(d) we plot the average mean and standard deviation of the inferred distribution $\PP (\beta)$, with the average taken over the different experiments. This shows that consistent accuracy of a few percent can be obtained even as $\beta$ is reduced over two orders of magnitude. We also count the fraction of times in which the true $\beta$ value lies in each decile of the Bayesian prediction. Ideally this would be exactly $0.1$ for each decile and our predictions conform to this, staying between $0.8$ and $1.2$. Plots of these values are given in the Supplemental Material.


\begin{figure}[tb]
        \begin{subfigure}[b]{1.68 in}
                \includegraphics[width=\textwidth]{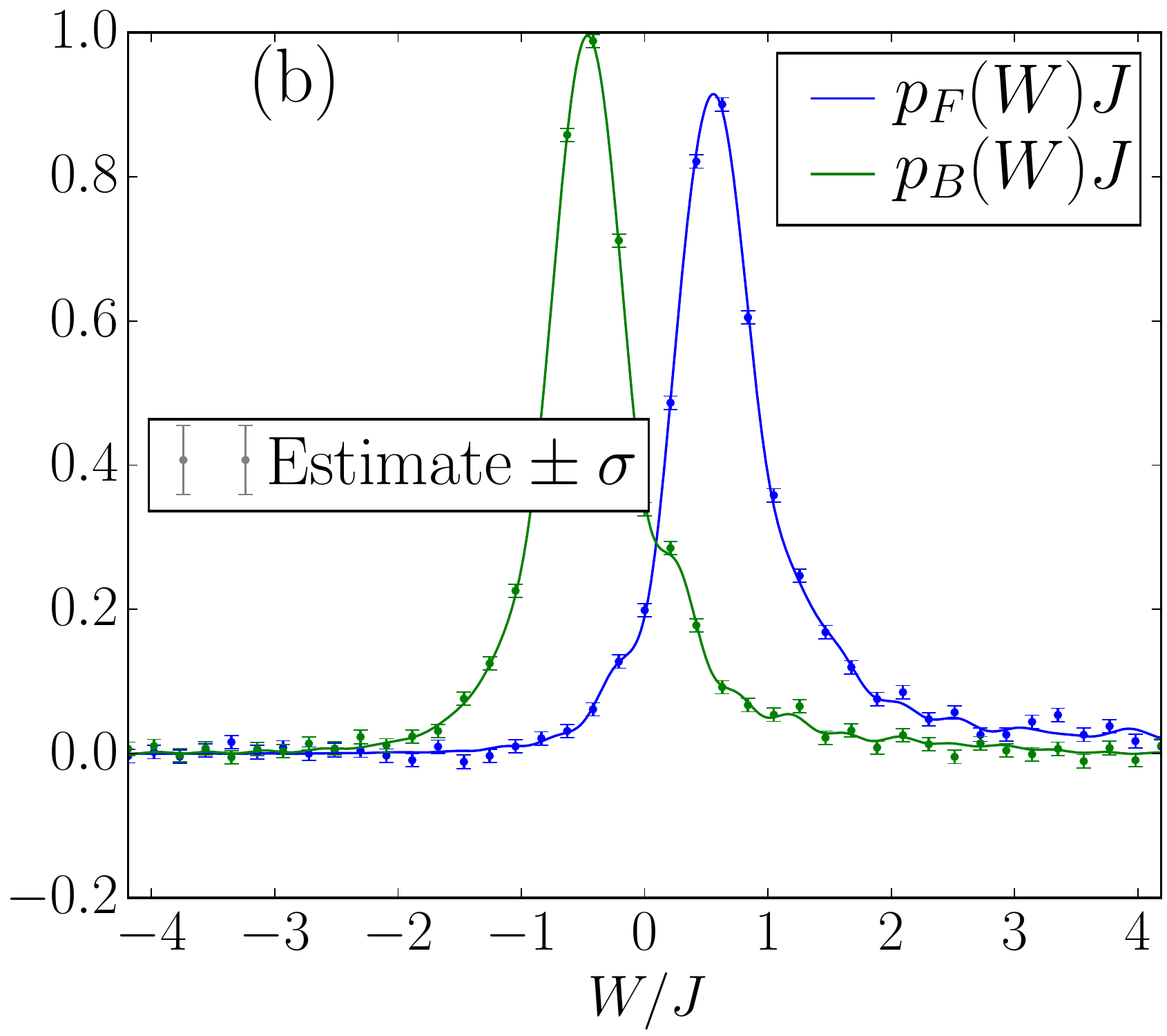}
                \label{fig:PsvsW}
        \end{subfigure}
        \begin{subfigure}[b]{1.68 in}
                \includegraphics[width=\textwidth]{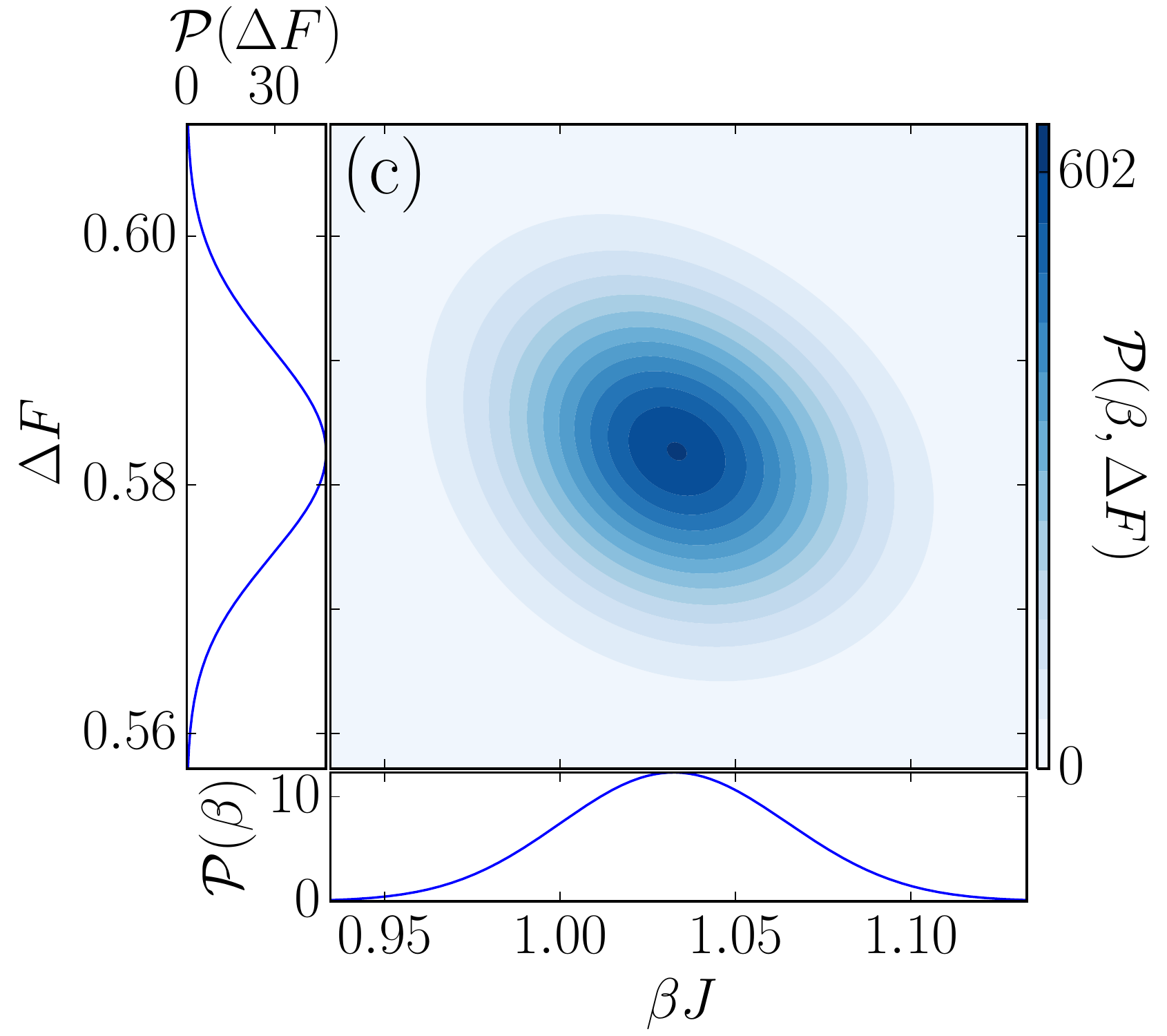}
                \label{fig:PsRatio}
        \end{subfigure}
        \caption{\label{fig:bhmtypicaltemp}{\em Stronger interactions}. The contents and parameters of this figure are identical to \fir{fig:typicaltemp}(b) and (c), but for a smaller $M=11$ system, featuring stronger interactions $U/J = 4$ and $\lambda_f \eta / J = 2$, and a shorter $\tau J = 0.1$ quench and longer window-size $T J = 15$.
        }
\end{figure}


\subsection{Stronger interactions} 

We now study the success of the protocol when the bosons are more strongly interacting. In this case the Bogoliubov approach is not valid, and instead our analysis proceeds using time-evolving block decimation~\cite{Vidal2004prl,White2004prl,Johnson2010pre} to evolve a matrix product operator~\cite{Verstraete2004prl,Zwolak2004prl,Schollwock2011aop} representation of the state of the bosons. Using this we near-exactly calculate the characteristic function $\chi_Q(u)$ for the exact Bose-Hubbard model [\eqr{BoseHubbardHamiltonian}] and interaction Hamiltonian. See the Supplemental Material for details on this tensor network method~\cite{TNTLibrary}.

The results for strongly-interacting bosons, close to the critical point, are shown in \fir{fig:bhmtypicaltemp}. Compared to superfluid bosons, we see that, despite stronger impurity-boson coupling, the qubit dephasing takes place over a longer timescale due to the absence of a broad spectrum of low-energy excitations. Correspondingly the work distributions $p_Q(W)$, shown in \fir{fig:bhmtypicaltemp}(a), are more featured than for the superfluid case, with positive skewness resulting from a high-frequency shoulder. However, as shown in \fir{fig:bhmtypicaltemp}(b), the accuracy of the thermometry procedure remains largely indifferent to these changes.

\section{Discussion}

We have shown that using a non-equilibrium probe overcomes two challenges in the thermometry of ultracold gases. First, the need to precisely control the internal energy of the probe on scales corresponding to the thermal energy of the system. Second, the need to understand the temperature-dependence of the system's thermal state in advance. To demonstrate this, we showed that the temperature of bosons in a lattice could be estimated using the protocol, for temperatures in the nanokelvins or even lower. We also found that accuracy and precision are largely unaffected by moving close to a critical point, in this case the crossover between superfluid and insulating phases, a regime that is less well understood.

These advantages come at a cost, namely the need for exquisite control over the interaction between the probe qubit and the cold-atom system. Furthermore, it must be possible to perform a quench such that the non-equilibrium work fluctuations are comparable to the thermal energy of the system. We have argued that atomic impurity probes can be expected to satisfy the aforementioned requirements quite generally, and are therefore excellent candidates for generic thermometry of cold atoms at very low temperatures.

Instead of performing repeated measurements on a single qubit probe, multiple measurements could be performed simultaneously using multiple probes. These could be prepared in an array using a lattice and used either to reduce the number of measurements per probe or to probe a spatially varying temperature profile resulting from e.g.\ heat currents during non-equilbrium transport~\cite{Brantut2013,Mendoza2013}. Alternatively, if the qubit probe is implemented by interfering atoms passing through the system at shifted times, then a steady stream of atom probes could near-continuously monitor the temperature of the system. 

An essential assumption underlying our thermometry protocol is that the system is in thermal equilibrium. However, the protocol could potentially be used to assess whether this is the case. The protocol is used to find the most likely pairs $\beta$ and $\Delta F$ given that the Tasaki-Crooks relation holds, but it could also evaluate the likelihood that the relation is satisfied for any $\beta$ and $\Delta F$, thus allowing the testing of thermalization~\cite{Eisert2015}.

\acknowledgments{The authors thank Sarah Al-Assam for her valuable assistance with the tensor network theory library~\cite{TNTLibrary}. THJ and DJ thank the National Research Foundation and the Ministry of Education of Singapore for support. MTM acknowledges financial support from EPSRC via the Controlled Quantum Dynamics CDT. We gratefully acknowledge financial support from the Oxford Martin School Programme on Bio-Inspired Quantum Technologies, the European Research Council under the European Union's Seventh Framework Programme (FP7/2007-2013)/ERC Grant Agreement no.\ 319286 Q-MAC and the collaborative project QuProCS (Grant Agreement 641277), and the EPSRC through projects EP/K038311/1 and EP/J010529/1. Data produced by EPSRC funded work is contained in the source code associated to the arXiv submission of this article.}
 
\bibliography{paper}

\newpage

\section*{Supplemental material}

This Supplemental Material contains some additional details relating to the calculations presented in the main text and is organized in the following way. We briefly introduce the relationship between the characteristic function and moments of the work distribution in \secr{sec:workdistprop}. Then, in \secr{sec:workdistcalc}, we give details regarding how estimates of the work distributions are obtained from estimates of the characteristic function, together with properties of these estimates. Section \ref{sec:infertemp} concerns how these estimates are used to infer the temperature, in both a frequentist and Bayesian approach. Then finally in \secr{sec:chicalc} we explain how the characteristic function is calculated for the specific case of the Bose-Hubbard model, using Bogoliubov theory and tensor network theory.

\section{Properties of the work distribution}
\label{sec:workdistprop}
In what follows we will refer to a few generic properties of the work distribution $P_Q(W)$, namely its cumulants $\kappa_m$ and principally its first $\kappa_{Q1} = \mu_Q$ and second $\kappa_{Q2} = \sigma_Q^2$ cumulants. The cumulants $\kappa_{Q m}$ of the work distribution $P_Q(W)$ are related to the derivatives of the logarithm of the corresponding characteristic function $\chi_Q (u)$, according to
\begin{align}
\label{eq:cumulants}
\kappa_{Qm} = & \frac{1}{\ii^m} \left. \frac{\dd^m}{\dd u^m} \log \chi_Q (u) \right |_{u=0} .
\end{align}

We note two properties of the work distribution $P_Q(W)$, one relevant to each of $\mu_Q$ and $\sigma_Q^2$. First, the second law of thermodynamics expressed in terms of mean work and free energy ensures that $\mu_F \geq \Delta F \leq -\mu_B$, where $\Delta F = F(\lambda_f) - F(\lambda_i)$ is the free energy difference. Second, though this is just a re-expression of \eqr{eq:cumulants} for $m=2$, the second cumulant is related to the so-called dephasing time $\tau_{Q\mathrm{deph}}$ by $\sigma_Q = 1 / \tau_{Q\mathrm{deph}}$, where $\tau_{Q\mathrm{deph}}$ is here defined by the short-time behavior of the dephasing function $|\chi_Q(u)| = \exp(-\Gamma_Q(u))$, with $\Gamma_Q (u) = (u/\tau_{Q\mathrm{deph}})^2/2 + \OO (u^3)$. This gives us a way of assessing the size of the second cumulant $\sigma_Q^2$ from observing a small part of $\chi_Q (u)$ directly. We assume that this is done, at a small overhead to our thermometry protocol discussed in the next section, and thus $\sigma_Q$ and $\tau_{Q\mathrm{deph}}$ are quantities that are, at least approximately, known when implementing the protocol.

\section{Estimating the work distribution}
\label{sec:workdistcalc}
\subsection{Deterministic errors}
Here we discuss how to estimate the work distribution
\begin{align}
P_Q (W) = \mathcal{F} \{ \chi_Q (u) \} (W) = \frac{1}{2 \pi} \int \dd u \ee^{- \ii W u} \chi_Q (u) , \nonumber
\end{align}
from estimates of the characteristic function $\chi_Q(u)$ obtained in a (numerical or actual) experiment, where $\mathcal{F}$ denotes a Fourier transform.

It is infeasible to study the characteristic function $\chi_Q(u)$ continuously over all times. More realistically the characteristic function $\chi_Q (u_j)$ is studied at a finite number $N_\mathrm{steps}$ of discrete times $u_j = j \Delta u$ for integer $j=1,\dots,N_\mathrm{steps}$ in some domain $[-T, T]$ with $T = N_\mathrm{steps} \Delta u$. Noting that $\chi_Q(0) = 1$ and $\chi_Q(u_j) = \chi^\ast_Q(-u_j)$, we then, instead of $P_Q (W)$, construct a discrete and finite-time Fourier transform
\begin{align}
p_Q (W) =& \frac{\Delta u }{2 \pi}\sum_{j = -N_\mathrm{steps}}^{N_\mathrm{steps}} \ee^{- \ii W u_j} \chi_Q (u_j) w(u_j) \nonumber \\
=& \frac{\Delta u }{2\pi} \left( 1 + 2 \Re \left \{\sum_{j = 1}^{N_\mathrm{steps}} \ee^{- \ii W u_j} \chi_Q (u_j) w(u_j)  \right \} \right), \nonumber
\end{align}
where we have introduced a windowing function $w(u)$.

As we will now discuss, $p_Q(W)$ differs significantly from $P_Q (W)$. However, the forward and backward distributions only enter into our thermometry protocol through their ratio. We show below that the ratios $p_F(W)/ p_B(-W)$ and $R(W) = P_F(W)/ P_B(-W)$ can be made identical, meaning we are able to work with the alternative distributions $p_Q(W)$ rather than $P_Q (W)$. For the rest of this subsection, we consider how to choose $T$ and $N_\mathrm{steps}$ for our protocol such that this is the case.

The two Fourier transforms $p_Q (W)$ and $P_Q (W)$ are related, using the Poisson summation formula and the convolution theorem, by
\begin{equation}
p_Q (W) = \left \{\mathcal{F} \{ w(u) \} (W) \star \sum_{k=-\infty}^{\infty} P_Q(W + k \Delta W) \right \} (W) , \nonumber
\end{equation}
where $\Delta W = 2 \pi / \Delta u$ and $\star$ represents a convolution. This demonstrates the two sources of error in going from $P_Q (W)$ to $p_Q (W)$. First, aliasing, where frequencies differing by $\Delta W$ cannot be distinguished by looking at a function at a discretized set of points. Second, spectral leakage, where contributions from one frequency leak to those nearby on a scale $\pi/T$ due to the finite resolution offered by the window of finite size $T$. The former leads to the sum and the latter to the convolution.

For the effects of aliasing to be small, we must have that $\sigma_Q / \Delta W \ll 1$ or $\sigma_Q \Delta u \ll 2 \pi$ so that the widths $\sigma_Q$ of the work distributions is much smaller than the periodicity of its approximation $p_Q(W)$. Another way to write this, in terms of the dephasing time $\tau_{Q\mathrm{deph}} = 1 / \sigma_Q$ is $\Delta u / \tau_{Q\mathrm{deph}} \ll 2 \pi$ or $N_{\mathrm{steps}} \gg T / \tau_{Q\mathrm{deph}}$. In all examples used in our work, the number of time-steps $N_{\mathrm{steps}}$ is easily large enough to ensure the effects of aliasing are negligible. 

The effects of spectral leakage on $p_Q (W)$ will always be significant for a finite system with a discrete set of energy levels. Importantly for our protocol, the same is not true for their ratios, as we now show. Consider aliasing to have a negligible effect, then
\begin{align}
\frac{p_F(W) }{ p_B(-W)} =&  \frac{\left \{\mathcal{F} \{ w(u) \} (W) \star P_F(W) \right \} (W)}{\left \{\mathcal{F} \{ w(u) \} (W) \star P_B(-W) \right \} (W)} \nonumber \\
=&  \frac{\left \{\mathcal{F} \{ w(u) \} (W) \star P_B(-W) R(W) \right \} (W)}{\left \{\mathcal{F} \{ w(u) \} (W) \star P_B(-W) \right \} (W)} \nonumber \\
\approx &  \frac{\left \{\mathcal{F} \{ w(u) \} (W) \star P_B(-W) \right \} (W) R(W) }{\left \{\mathcal{F} \{ w(u) \} (W) \star P_B(-W) \right \} (W)} \nonumber \\
= & R(W) . \nonumber
\end{align}
In going from the second to the third line we have only used that $R(W) = \ee^{\beta (W-\Delta F)}$ does not vary too much on the scale of the characteristic width $\Delta W_{\mathcal{F}w} \approx \pi / T$ of the smoothing kernel $\mathcal{F} \{ w(u) \} (W)$, i.e., $\beta \Delta W_{\mathcal{F}w} \ll 1$ or $T \gg \pi \beta$. However, even this criterion is sometimes overly strict and may be loosened depending on the relationship between $\sigma_Q$, $\beta$ and $\Delta W_{\mathcal{F}w}$. For example, consider the case that both $p_Q(W)$ are effectively flat on scale $\Delta W_{\mathcal{F}w}$ i.e.\ $\sigma_Q / \Delta W_{\mathcal{F}w} \gg 1$ assuming $p_Q(W)$ to be unimodal. Then even having $\beta \Delta W_{\mathcal{F}w} \approx 1$ would lead only to small errors, as errors due to variations in $R(W)$ within the convolution would largely cancel. Note that if an incorrect choice is made and spectral leakage does introduce errors, then this will be clear from the non-linear behavior of $L(W) = \ln (R(W))$ and thus a larger $T$ can be chosen.

In this work, we find that a good rule of thumb is to choose $T$ to be sufficiently long that the qubit has fully dephased. Specifically, for the superfluid calculations, we use a phenomenological and unoptimized expression based on the above discussion
\begin{align}
\label{eq:Tchoice}
T = \pi \beta [1 + (5 - 1) \ee^{-\sigma \beta} ] ,
\end{align}
with $\sigma = \sigma_F = \sigma_B$. The effect is that spectral leakage, like aliasing, has a negligible effect on errors. In a real experiment, $\beta$ is not known in advance and so $T$ must be chosen using the considerations above and some prior expectations about $\beta$. We find that the thermometry protocol is typically robust to changes in $T$ by factors of the order of unity.

\subsection{Random errors}
From now on, we assume that $T$ and $N_{\mathrm{steps}}$ have been chosen, such that $R(W) = p_F(W) / p_B(-W)$. We focus on the fact that $\chi_Q(u_j)$ will not be known exactly and that instead we only have access to an estimator $\bar{\chi}_Q (u)$ based on expectation values estimated from a finite number $N_\mathrm{meas}$ of measurements each. We will always use a bar to indicate an estimate that is a random variable obtained stochastically from measurements made during the protocol. Propagating this forward according to
\begin{align}
\label{eq:ftestimates}
\bar{p}_{Q}(W) = \frac{\Delta u }{2\pi} \left( 1 + 2 \Re \left \{\sum_{j = 1}^{N_\mathrm{steps}} \ee^{- \ii W u_j} \bar{\chi}_{Q} (u_j) w(u_j)  \right \} \right),
\end{align}
we obtain an estimate $\bar{p}_{Q}(W)$ of $p_Q(W)$. The remainder of this subsection addresses how to choose a set of work values $W_k$ such that $\bar{p}_{Q} (W_k)$ are independent, unbiased $\EE [ \bar{p}_{Q} (W_k)] = p_Q(W_k)$, and of known variance $\Var[\bar{p}_{Q} (W_k)]$. Here expectation values are always taken with respect to the distributions generating the measurement outcomes.

Let us begin by considering how $\chi_Q(u_j)$ are estimated. In an experiment we estimate
\begin{align}
\chi_Q (u) =  \frac{\ee^{\ii \phi(u)}}{2 s^\ast_\uparrow s_\downarrow} \left( \langle \hat{\sigma}_x \rangle + \ii \langle \hat{\sigma}_y \rangle \right) , \nonumber
\end{align}
with $\phi(u) = \Delta (\tau + u)$ by first estimating expectation values $\langle \hat{\sigma}_\mu \rangle$, for $\mu = x,y$, with respect to the state $\hat{\rho}_{q}$ of the qubit at the end of the interferometric protocol.
Specifically, each $\langle \hat{\sigma}_\mu \rangle$ is estimated from $N_\mathrm{meas}$ independent measurements of $\sigma_\mu$, which returns $1$ or $-1$ with probability $p = (1 + \langle \hat{\sigma}_\mu \rangle)/2$ and $1-p$, respectively. Then the average $\bar{\sigma}_\mu$ of the measurements is an estimator of $\langle \hat{\sigma}_\mu \rangle$ that is unbiased, i.e., its mean is $\EE [\bar{\sigma}_\mu ] =  \langle \hat{\sigma}_\mu \rangle$, and has variance $\Var [\bar{\sigma}_\mu ] = (1- \langle \hat{\sigma}_\mu \rangle^2) / N_\mathrm{meas}$. 
In accordance with the central limit theorem, for large enough $N_\mathrm{meas}$ the estimator $\bar{\sigma}_\mu$ is normal and so its properties are fully characterized by its mean and variance. 

The linear combination 
\begin{align}
\bar{\chi}_Q (u) =  \frac{\ee^{\ii \phi (u)}}{2 s^\ast_\uparrow s_\downarrow} \left( \bar{\sigma}_x + \ii \bar{\sigma}_y \right) , \nonumber
\end{align}
is thus also an unbiased ($\EE [\bar{\chi}_Q (u) ] =  \chi_Q (u)$) Gaussian estimator of $\chi_Q (u)$ and has variance (using the generalized definition $\Var [\bar{z}] = \EE \left [ |\bar{z}- \EE [\bar{z}]|^2  \right ]$ of variance for a complex random variable $\bar{z}$)
\begin{equation}
\label{eq:chivar}
\begin{aligned}
\Var [\bar{\chi}_Q (u)  ] = & \frac{\Var [\bar{\sigma}_x ] + \Var [\bar{\sigma}_y ] }{ 4 | s^\ast_\uparrow s_\downarrow |^2} \\
 = & \frac{ 2 - 4 | s^\ast_\uparrow s_\downarrow |^2 |\chi_Q (u)|^2 }{ 4  | s^\ast_\uparrow s_\downarrow |^2 N_\mathrm{meas} } , 
\end{aligned}
\end{equation}
where the first line holds due to the independence of $\bar{\sigma}_x$ and $\bar{\sigma}_y$.

Note that this variance $\Var [\bar{\chi}_Q (u)  ] = \Var [ \Re \{ \bar{\chi}_Q (u) \}  ] + \Var [ \Im \{ \bar{\chi}_Q (u) \}   ]$ is divided between those of the real and complex parts, but how exactly this division occurs depends on the phase $\phi(u) =  \Delta (\tau + u)$, and thus $\Delta$, and the phase of $s^\ast_\uparrow s_\downarrow$. The effect of $\Delta$ is quantitative but not qualitative, and the $\Delta$ for a particular implementation can be known. For the purposes of our plots, we assume $\phi(u) = \Delta = 0$. Further, \eqr{eq:chivar} makes it obvious that $s^\ast_\downarrow$ and $s^\ast_\uparrow$ should be chosen such that $| s^\ast_\uparrow s_\downarrow |^2$ takes its maximum possible value, $1/4$, and in all plots we assume the choice $s^\ast_\uparrow s_\downarrow = 1/2$ is made.

This variance propagates forward into our estimator $\bar{p}_{Q}(W)$ of $p_Q (W)$ according to \eqr{eq:ftestimates}. Since this again a linear sum, the estimators are unbiased and Gaussian, characterized by the expectations $\EE [\bar{p}_{Q}(W)] = p_Q (W)$ and covariances
\begin{widetext}
\begin{equation}
\label{eq:pcovar}
\begin{aligned}
 \EE \big [ & \left (\bar{p}_{Q}(W) - p_Q (W) \right)  \left (\bar{p}_{Q}(W') - p_Q (W') \right) \big ] 
\\ = & \frac{2}{4 | s^\ast_\uparrow s_\downarrow |^2 N_\mathrm{meas} }  \left(  \frac{\Delta u }{2\pi} \right)^2 \sum_j   |w (u_j)|^2 \cos \left((W - W') u_j\right) (2 - 4 | s^\ast_\uparrow s_\downarrow |^2 |\chi_Q (u_j)|^2 )  \\
&- \frac{2}{4 (s^\ast_\uparrow s_\downarrow )^2 N_\mathrm{meas} }  \left(  \frac{\Delta u }{2\pi} \right)^2 \sum_j  w^2 (u_j) \Re \left \{ \ee^{-\ii (W+W') u_j} \ee^{2\ii \phi (u_j)} \Re \left \{ \ee^{-2\ii \phi (u_j)} 4 (s^\ast_\uparrow s_\downarrow )^2 \chi^2_Q (u_j) \right \} \right \}  .
\end{aligned}
\end{equation}
\end{widetext}
The first term arises from the interference of random deviations in $\chi_Q (u_j)$ at different $u_j$.
The second term, which is much smaller, arises due to the fact that, to ensure that $\bar{p}_{Q}(W)$ is real, we have used $\chi_Q (-u_j) = \chi^\ast_Q (u_j)$ rather than generate independent estimators for $\chi_Q (u_j)$ and $\chi_Q (-u_j)$. 

When $W$ and $W'$ are close, approximating $W = W'$, we obtain the variance. The dominant term is given by 
\begin{equation}
\label{eq:pvar}
\begin{aligned}
\Var \left [ \bar{p}_{Q}(W)  \right ] \approx& \frac{2}{4 | s^\ast_\uparrow s_\downarrow |^2 N_\mathrm{meas} } \left( \frac{\Delta u }{2\pi} \right)^2 \sum_j |w (u_j)|^2 \\ &\times
\Bigg ( 2 - 4 | s^\ast_\uparrow s_\downarrow |^2 |\chi_Q (u_j)|^2  \Bigg ) .
\end{aligned}
\end{equation}
Notably, this scales as $\Var \left [ \bar{p}_{Q}(W)  \right ] \sim  T^2 / N_\mathrm{steps} N_\mathrm{meas}$. Compare this to the conditions $T / \tau_{Q \mathrm{deph}} \ll 2 \pi N_\mathrm{steps}$ and $T \gtrsim \pi \beta$ for avoiding systematic errors due to aliasing and spectral leakage. The form of $\Var \left [ \bar{p}_{Q}(W)  \right ]$ suggests increasing $N_\mathrm{steps}$, which would also reduce the aliasing error. However, $\Var \left [ \bar{p}_{Q}(W)  \right ]$ suggests taking a smaller window size $T$, which would come at the cost of higher spectral leakage. We leave a discussion of the trade-off of these two errors until later.

When $W$ and $W'$ are separated by more than roughly $\pi/T$, the cosine term in \eqr{eq:pcovar} will oscillate rapidly enough that the covariance is significantly reduced. Thus $\pi/T$ represents the range in $W$ over which our estimates $\bar{p}_{Q}(W)$ are correlated. Thus there is little information added by generating estimates $\bar{p}_{Q}(W)$ for $W$ separated by less than this correlation range $\pi/T$. As a result, in our inference protocol we consider only the values $\bar{p}_{Q}(W_k)$ taken at an array of points $W_k = k \pi/T$ separated by this amount. These values should only be weakly-correlated. Conveniently, but not crucially, these are exactly the same values $W_k$ for which $\bar{p}_Q (W_k)$ can be found using the fast Fourier transform. This discussion suggests increasing $T$ in order to increase the density of points $W_k$. Again, a discussion of the trade-off of this with the other factors affecting the choice of $T$ is left until later.

\section{Inferring the temperature}
\label{sec:infertemp}
Here we detail the core of the thermometry protocol, giving the precise procedure to go from the estimates of the work distributions $p_Q(W_k)$ discussed in the previous section to an estimate of the inverse temperature $\beta$. 
Note that the protocol uses no knowledge specific about the distributions $P_Q(W)$ and characteristic functions $\chi_Q (u)$ other than the width $\sigma_Q$ and dephasing time $\tau_{Q\mathrm{deph}}$ described above.

\subsection{Outline}
The essential fact on which we base our inference is that
\begin{equation}
\label{eq:CrooksTasaki}
L(W) = \ln( R(W) ) = \ln \left \{ \frac{p_F(W)}{p_B(-W)} \right \} = \beta(W-\Delta F) ,
\end{equation}
where we refer to the work distributions $p_Q(W)$, ratio $R(W)$ and log-ratio $L(W)$ discussed in the previous section.

The information we collect during our protocol is a set of near-independent estimates $\bar{p}_Q (W_k)$ of $p_Q (W_k)$ and their variances at a discrete set of energies $W_k$. The essential idea is to use these estimated values of the work distribution, together with our knowledge of how they relate to $\beta$ and $\Delta F$ via \eqr{eq:CrooksTasaki}, to infer the values, or distribution of values, of $\beta$ (and $\Delta F$ if desired). 

We consider two approaches to this inference problem, frequentist and Bayesian. The Bayesian approach is used for all results we present in the main text, but methods along the lines of the frequentist approach are more commonly found in the literature. Our presentation of the frequentist approach thus serves to highlight the benefits our Bayesian approach, and also provides a simpler setting in which to analyze the dependence of errors on parameters e.g.\ $T$ used in the protocol.

\subsubsection{Frequentist}
The frequentist approach is most similar to how $\Delta F$ has previously been estimated from work distributions. The approach is to use $\bar{p}_Q (W_k)$ to obtain estimates $\bar{L}(W_k)$ of $L(W_k)$ for many values of work $W_k$. For now, let's assume, these estimates $\bar{L}(W_k)$ are independent, unbiased $\EE[\bar{L}(W_k)] = L(W_k)$, and Gaussian with known variance $\Var[\bar{L}(W_k)]$. Assuming this, the knowledge of \eqr{eq:CrooksTasaki} means that we can use weighted linear regression to construct the maximally likely $\bar{\beta}$ and $\overline{\Delta F}$ as the pair of values that minimize the least-squared error
\begin{align}
\label{eq:lserror}
\sum_k \frac{( \bar{L}(W_k) - \beta(W_k-\Delta F) )^2}{\Var[ \bar{L} (W_k) ]} , 
\end{align}
and take them as our estimates of $\beta$ and $\Delta F$. This is shown in \fir{fig:typicaltempfreq}. The procedure then boils down to obtaining estimates $\bar{L}(W_k)$ whilst ensuring no bias and estimating $\Var[\bar{L}(W_k)]$.

As a starting point, consider the estimator $\bar{L}' (W) = \ln \{ \bar{p}_F (W) / \bar{p}_B (-W)  \}$. Due to the non-linear nature of the inverse and logarithm functions, $\bar{L}' (W)$ is neither Gaussian nor an unbiased estimator of $L(W)$. The errors resulting from the bias can be small, but ideally we would like to construct an unbiased estimator $\bar{L} (W) = \bar{L}' (W) - \Delta L(W)$, which removes the bias $\Delta L(W) =\EE [ \bar{L}' (W) ] - L(W)$. We would also like to characterize the remaining mean-zero random errors in $\bar{L} (W)$ and find when they are approximately Gaussian. 

To do this, we approximate the bias and variance by simulating the sampling of $\bar{p}_Q(W)$, by adding Gaussian noise of zero mean and variance $\Var [ \bar{p}_Q (W)]$ to $\bar{p}_Q(W)$ before taking the logarithm. From these values, we then estimate the bias $\Delta L(W)$ and variance $\Var [ \bar{L} (W)]$, and assess the non-Gaussian character of the distribution of $\bar{L}(W)$. As is expected, we find $\bar{L}(W)$ is reasonably Gaussian when $\Var [ \bar{p}_Q(W) ]/p^2_Q(W)$ is small. For larger values the distribution is very skewed and least-squares minimization may not correspond to the maximally likely parameters. 

A further flaw in this approach is the possibility of obtaining negative values $\bar{p}_Q(W)$, whose logarithm is undefined. We simply ignore values of $W_k$ for which $\bar{p}_F(W_k)$ or $\bar{p}_B(-W_k)$ is negative, and we ignore negative values arising in our estimate of the bias and variance. This comes at a cost of potentially inducing a bias. Balancing the need to avoid such a bias with the desire to have as many points as possible for the fit, in our frequentist calculations presented here we only consider values $W_k$ where $\Var [ \bar{p}_F( W_k) ] / p^2_F(W_k) > 1$ and $\Var [ \bar{p}_B( -W_k) ] / p^2_B(-W_k) > 1$. The results in \fir{fig:lowtempfreq} show that this bias is perhaps acceptable but not small. A better approach for dealing with negative values is the Bayesian approach of the next subsection, which makes full use of the information provided by obtaining a negative value for $\bar{p}_F(W_k)$ or $\bar{p}_B(-W_k)$. For now, we continue our analysis of the frequentist approach ignoring any bias introduced by these negative values.

We have shown how to choose a set of $W_k$ for which we have approximately independent, unbiased, and Gaussian random estimators $\bar{L}(W_k)$ of $L(W_k)$ whose variances $\Var[ \bar{L} (W_k) ]$ we know approximately. From these, we then obtain a maximally likely $\bar{\beta}$ and $\overline{\Delta F}$ as the pair of values that minimize the least-squared error of \eqr{eq:lserror} and take them as our estimates of $\beta$ and $\Delta F$. Let us now turn to discussing how the variance in $\bar{\beta}$ should behave, including how it depends on some of the parameter choices we make in our protocol, particularly $T$, since the dependence of the error on $N_{\mathrm{meas}}$ and $N_{\mathrm{steps}}$ is clear.

It is well known that, for least-squares estimation, the fitted parameter $\bar{\beta}$ provides an unbiased estimate of $\beta$ with variance
\begin{align}
\Var [ \bar{\beta} ] =& \frac{1/\varsigma_1}{\varsigma_{W^2}/\varsigma_1 - (\varsigma_{W}/\varsigma_1)^2} , \nonumber \\
\varsigma_1 =& \sum_k \frac{1}{\Var[ \bar{L} (W_k) ]} ,  \nonumber \\
\varsigma_{W} =& \sum_k \frac{W_k}{\Var[ \bar{L} (W_k) ]} ,  \nonumber \\
\varsigma_{W^2} =& \sum_k \frac{W_k^2}{\Var[ \bar{L} (W_k) ]} . \nonumber 
\end{align}
We obtain a simpler expression for the fractional error
\begin{align}
\frac{\sqrt{\Var [ \bar{\beta} ]}}{\beta} = & \frac{\sqrt{\Var[ \bar{L} ] /N_W}}{\beta \Delta W_k} , \nonumber 
\end{align}
that qualitatively captures the basic behavior if we assume uniform variance $\Var[ \bar{L} (W_k) ] \sim \Var[ \bar{L} ]$ with $\Var[ \bar{L} ]$ the typical value of the variance over the values of $W_k$ used.
Appearing in the denominator of this equation is the spread $\Delta W_k = \sqrt{\sum_k W_k^2/N_W - (\sum_k W_k/N_W)^2}$ of the values $W_k$, where we have written $N_W = \sum_k$ for the number of points used in the estimation.

We can now insert some of the findings from the error analysis of the previous sections into this expression. First, demanding independence of points required us to choose that $W_k$ were spaced by $\pi/T$ and so $N_W \sim \Delta W_k T /\pi$, giving us
\begin{align}
\frac{\sqrt{\Var [ \bar{\beta} ]}}{\beta} \sim & \frac{\sqrt{\Var[ \bar{L} ] /T}}{\beta (\Delta W_k)^{3/2}} .  \nonumber 
\end{align}
Second, the range of work values $\Delta W_k$ we include is the range satisfying $\Var [ \bar{p}_F( W_k) ] / p^2_F(W_k) > 1$ and $\Var [ \bar{p}_B( -W_k) ] / p^2_B(-W_k) > 1$, which corresponds to ensuring that $T^2 / N_{\mathrm{meas}} N_{\mathrm{steps}} p_F^2(W_k)$ and $T^2 / N_{\mathrm{meas}} N_{\mathrm{steps}} p_B^2(-W_k)$ are smaller than a threshold amount. This tells us that this range $\Delta W_k$ can decay quickly with increasing $T$ if $p_Q(W)$ depends sharply on $W$ at the edge of the region $W_k$. In turn this means that $T$ should be chosen to be as small as possible without introducing spectral leakage, which is the solution to one of the main questions remaining from the above discussion. We found these optimum values of $T$ to behave roughly as \eqr{eq:Tchoice}, essentially linearly in $\beta$, leaving us with
\begin{align}
\frac{\sqrt{\Var [ \bar{\beta} ]}}{\beta} \sim & \frac{\sqrt{\Var[ \bar{L} ]}}{ (\beta \Delta W_k)^{3/2}} .  \nonumber 
\end{align}
It is difficult to ascertain from this simplified analysis how this fractional error depends on $\beta$. In our examples we find that the relevant width $\Delta W_k$ of work values decreases roughly linearly with $\beta$ for large $\beta$. The variances captured by $\Var[ \bar{L} ]$ are the deciding factor. We find that in our examples they increase with $\beta^2$ as might be predicted from $\Var[ \bar{p}_Q (W) ] \sim T^2$ and our setting of $T \sim \beta$. Thus the fractional variance of our $\beta$ estimate increases roughly linearly with $\beta$ [\fir{fig:lowtempfreq}].


\begin{figure}[tb]
        \begin{subfigure}[b]{1.68 in}
                \includegraphics[width=\textwidth]{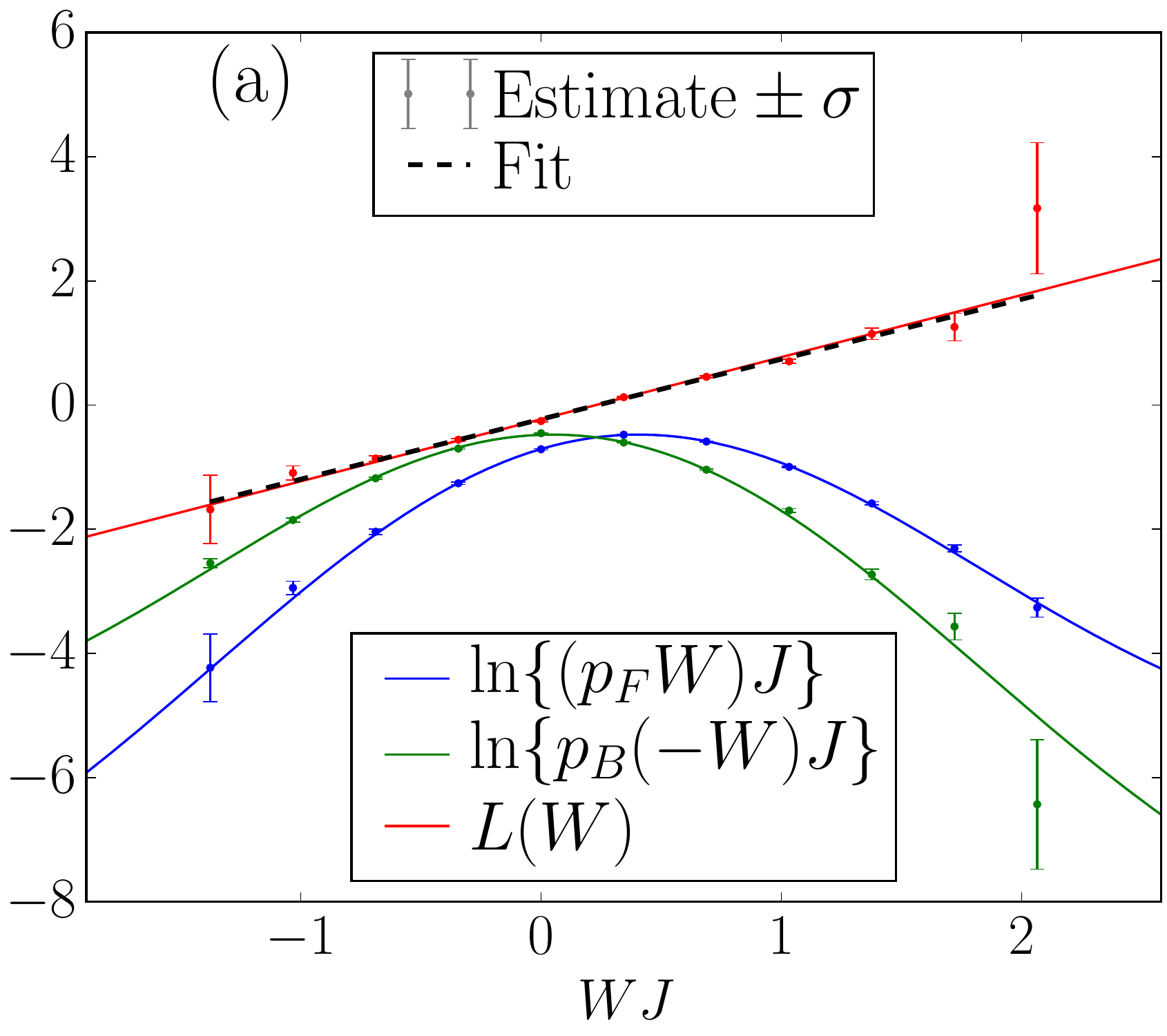}
                \label{fig:PsRatio}
        \end{subfigure}
        \begin{subfigure}[b]{1.68 in}
                \includegraphics[width=\textwidth]{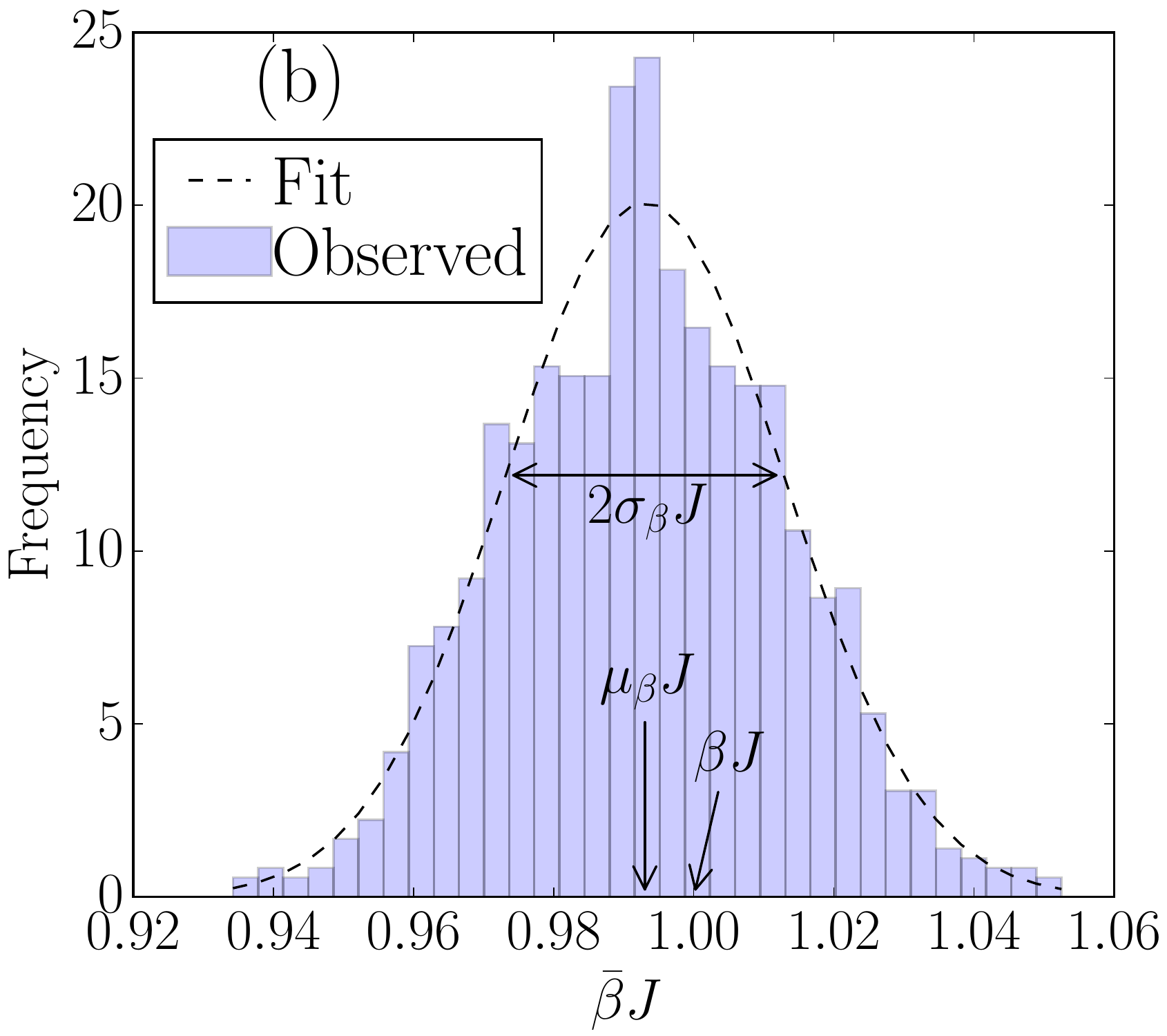}
                \label{fig:betahist}
        \end{subfigure}
        \vspace{-20pt}
        \caption{\label{fig:typicaltempfreq}{\em Superfluid phase; frequentist}. (a) The logarithms $\ln \{ p_F (W) \}$ and $\ln \{ p_B (-W) \}$ of the work distributions and their difference $L(W) = \ln \{ p_F (W) \} - \ln \{ p_B (-W) \}$. Solid lines are known values, points are estimates obtained in a single simulated experiment, and error bars calculated from those estimates mark expected errors of one standard deviation. The black dashed line is obtained from a weighted least squares fit. (b) A histogram of $\bar{\beta}$ estimates obtained in 1000 simulated experimental runs, together with their mean $\mu_\beta$ and standard deviation $\sigma_\beta$, and Gaussian fit. All parameters are identical to those for Fig.\ 1 in the main text.}
\end{figure}



\begin{figure}[tb]
\includegraphics[width=1.68 in]{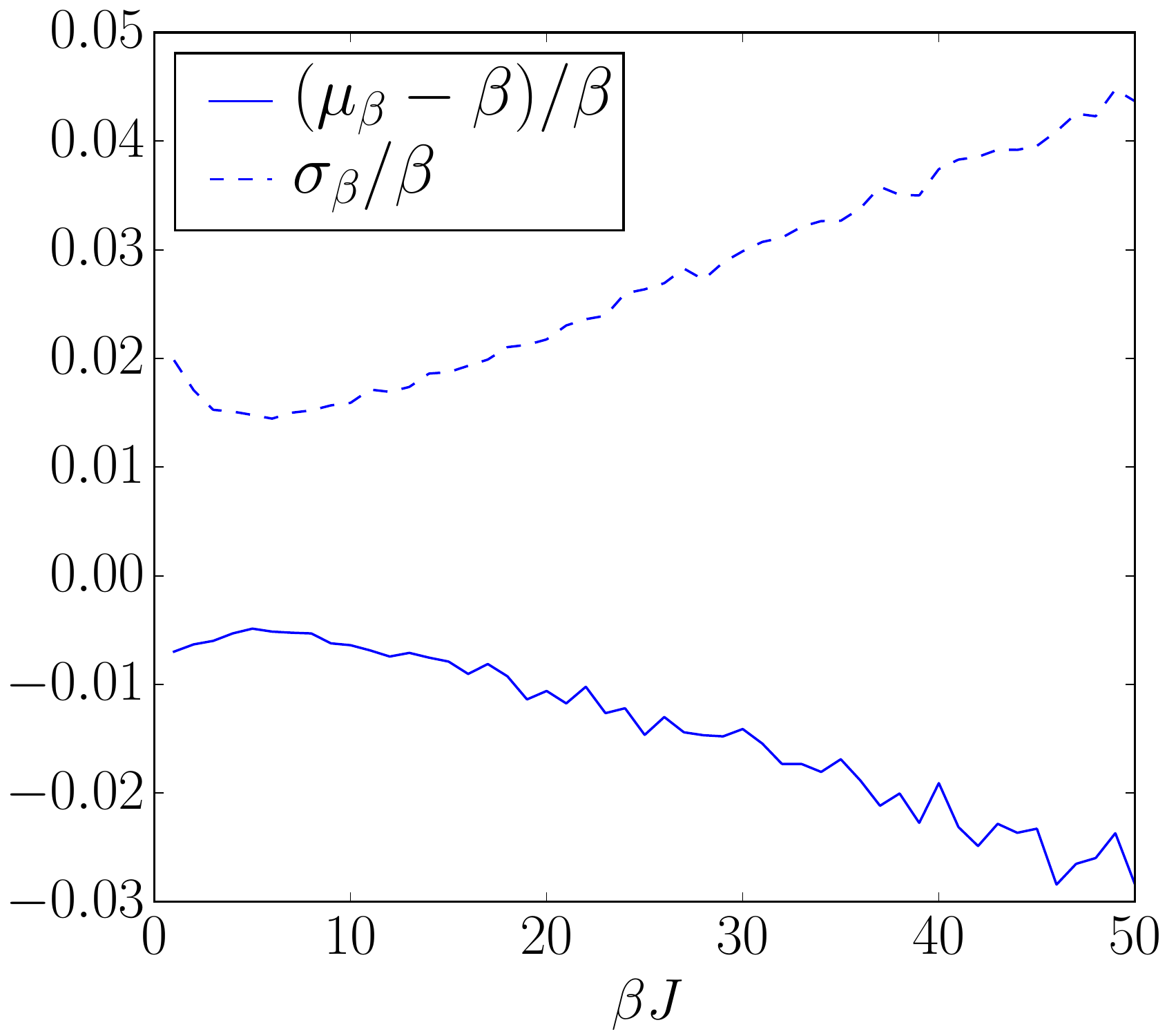}
\caption{\label{fig:lowtempfreq}{\em Accuracy at low temperatures; frequentist}. Over 1000 simulated experiments for each $\beta$, the average fractional standard deviation $\sigma_\beta /\beta$ and bias $(\mu_\beta - \beta) / \beta$. All other parameters are identical to those for Fig.\ 1(d) in the main text.}
\end{figure}


\subsubsection{Bayesian}
The Bayesian approach is to capture probabilistically the knowledge we have about $\beta$ and $\Delta F$ conditioned upon obtaining estimates $\bar{p}_Q(W)$. It also allows us to include our prior expectations about the system into the statistical analysis. However, we do not make use of that feature here, assuming nothing about the system. Further, the Bayesian approach outputs a probability distribution of the values of $\beta$ and $\Delta F$, which need not be Gaussian, rather than only returning maximally likely parameters and an estimate of their variance. The Bayesian approach therefore can use and provide more information than the frequentist approach.

In the following we use the shorthand notation $p_{F} = p_F (W_k)$, $p_{B} = p_B (-W_k)$, $R = R(W_k)$ and $W = W_k$. We write $O_k$ to denote the observations of estimates $\bar{p}_{F}$ and $\bar{p}_{B}$ for some value of $k$ and $O$ for the combined set of observations.

The Bayesian approach is based on the expression
\begin{align}
\label{eq:Bayeslawmain}
\PP(\beta, \Delta F | O) = \frac{ \PP( O | \beta, \Delta F) \PP(\beta, \Delta F)}{ \int \dd \beta \dd \Delta F \PP( O | \beta, \Delta F) \PP(\beta, \Delta F)  } ,
\end{align}
for our assessment of the probability of the system having temperature $\beta$ and free energy difference $\Delta F$ given observations $O$. It uses our assumptions about $\PP(O | \beta, \Delta F)$, the probability we would have obtained those observations for all possible values of $\beta$ and $\Delta F$, and the prior $\PP(\beta, \Delta F)$, capturing our knowledge of the system in advance of the experiment.

In this paper we make what is called a null prior, setting $\PP(\beta, \Delta F)$ to be constant, essentially assuming we have no information about $\beta$ and $\Delta F$. An experimentalist who does have more prior information could easily adapt our approach to include that information in the inference scheme.

We are left then to come up with an expression for the conditional probability $\PP( O| \beta, \Delta F)$, which, assuming independence, is just a product of the conditional probability of obtaining pairs of observations at each $W$. The next few paragraphs deal with the evaluation of this conditional probability. We perform this evaluation by breaking it up into several parts, in three stages. First we use that 
\begin{align}
\PP( \bar{p}_{F}, \bar{p}_{B} | \beta, \Delta F) &= \int \dd R \PP( \bar{p}_{F}, \bar{p}_{B} | R ) \PP( R | \beta, \Delta F) \nonumber \\
&= \PP( \bar{p}_{F}, \bar{p}_{B} | \ee^{\beta (W - \Delta F)} ) . \nonumber
\end{align}
Here we have conditioned the obtaining of the estimates $\bar{p}_F$ and $\bar{p}_B$ on the ratio $R$ of their true values, and used fact that this ratio is always equal to $\ee^{\beta (W - \Delta F)}$. It is this piece of information that is the key to the whole protocol.

The second step is to use
\begin{align}
\PP( \bar{p}_F, \bar{p}_B | R) &= \int \dd p_F \dd p_B \PP( \bar{p}_F, \bar{p}_B | p_F, p_B) \PP( p_F, p_B | R ) . \nonumber
\end{align}
Here we have conditioned on the exact values of the work distributions $p_Q$. We do this because we know that $\bar{p}_Q$ are unbiased and Gaussian with known variance $\sigma_Q$ i.e.\ $\PP( \bar{p}_F, \bar{p}_B | p_F, p_B ) = \PP( \bar{p}_F| p_F) \PP( \bar{p}_B| p_B)$ with
\begin{align}
\PP( \bar{p}_Q| p_Q) =& \frac{1}{\sqrt{2 \pi \sigma_Q^2}} \ee^{-(\bar{p}_Q - p_Q)^2/2 \sigma_Q^2} . \nonumber
\end{align}

The distribution $\PP( p_F, p_B | R)$ of the work distributions given their ratio can be built in our third and final step. We use Bayes' law again
\begin{align}
\PP( p_F, p_B | R) =\frac{ \PP( R | p_F, p_B) \PP(p_F, p_B)}{ \int \dd p_F \dd p_B \PP( R | p_F, p_B) \PP(p_F, p_B)  } , \nonumber
\end{align}
since we know that
\begin{align}
\PP( R | p_F, p_B) = \delta( R - p_F / p_B) = p_B \delta( R p_B - p_F)  . \nonumber
\end{align}
This leaves us having to define some prior $\PP(p_F, p_B)$ representing our prior knowledge of the work distribution values. We again essentially assume no prior knowledge, assuming independence $\PP(p_F, p_B) = \PP(p_F ) \PP ( p_B)$ and a uniform distribution over positive values
\begin{align}
\PP(p_Q ) = \begin{cases} \frac{1}{C} & \quad \text{if } 0 \leq p_Q < C \\ 0 & \quad \text{otherwise}\\ \end{cases} , \nonumber
\end{align}
for some $C$ taken to be much bigger than all other values.

We now have all we need. Collecting all of the above together, after performing two integrals, we obtain
\begin{align}
&\PP( p_F, p_B | R) = \nonumber \\
&\qquad 2 \max \{ 1, 1 +  R^2  \} p_B \delta( R p_B - p_F) \PP(p_F ) \PP ( p_B), \nonumber
\end{align}
and
\begin{widetext}
\begin{align}
&\PP( \bar{p}_F, \bar{p}_B | \beta, \Delta F) = \nonumber
\\ &  \qquad \frac{\max \{1, 1 +  R^2  \}}{2 \pi C^2 R} \left ( \exp \left [ -\frac{p_F^{\prime 2}}{2 \sigma_F^{\prime 2}} -\frac{p_B^{2}}{2 \sigma_B^{2}}  \right] \frac{\sigma_F^{\prime} \sigma_B}{\pi \sigma_{FB}^{\prime 2} } + \exp \left [ -\frac{(p_F^{\prime} - p_B)^2}{2 \sigma_{FB}^{\prime 2}}\right]  \frac{p_F^{\prime} \sigma_B^2 + p_B \sigma_F^{\prime 2}}{\sqrt{2 \pi} \sigma_{FB}^{\prime 3}} \left [ 1 + \mathrm{erf} \left( \frac{p_F^{\prime} \sigma_B^2 + p_B \sigma_F^{\prime 2}}{\sqrt{2} \sigma_F^{\prime} \sigma_B \sigma_{FB}^{\prime} } \right)  \right]   \right), \nonumber
\end{align}
\end{widetext}
where we have used the shorthand $p_F^{\prime} = p_F / R$, $\sigma_F^{\prime} = \sigma_F /R$ and $\sigma_{FB}^{\prime 2} = \sigma_F^{\prime 2} + \sigma_B^2$.

All of this can be fed back into our expression for $\PP (\beta, \Delta F|O)$ [\eqr{eq:Bayeslawmain}] and allows us to plot this distribution and calculate its properties. Upon testing, the Bayesian prediction is found to be remarkably well calibrated. We have performed $1000$ simulated experiments and the fraction of times in which the true $\beta$ value lies in each decile of the Bayesian prediction. Ideally this would be exactly $0.1$ for each decile and our predictions conform to this, staying between $0.8$ and $1.2$ for $\beta$ between $1$ and $50$, as shown in \fir{fig:calibration}.


\begin{figure}[tb]
\includegraphics[width=1.68 in]{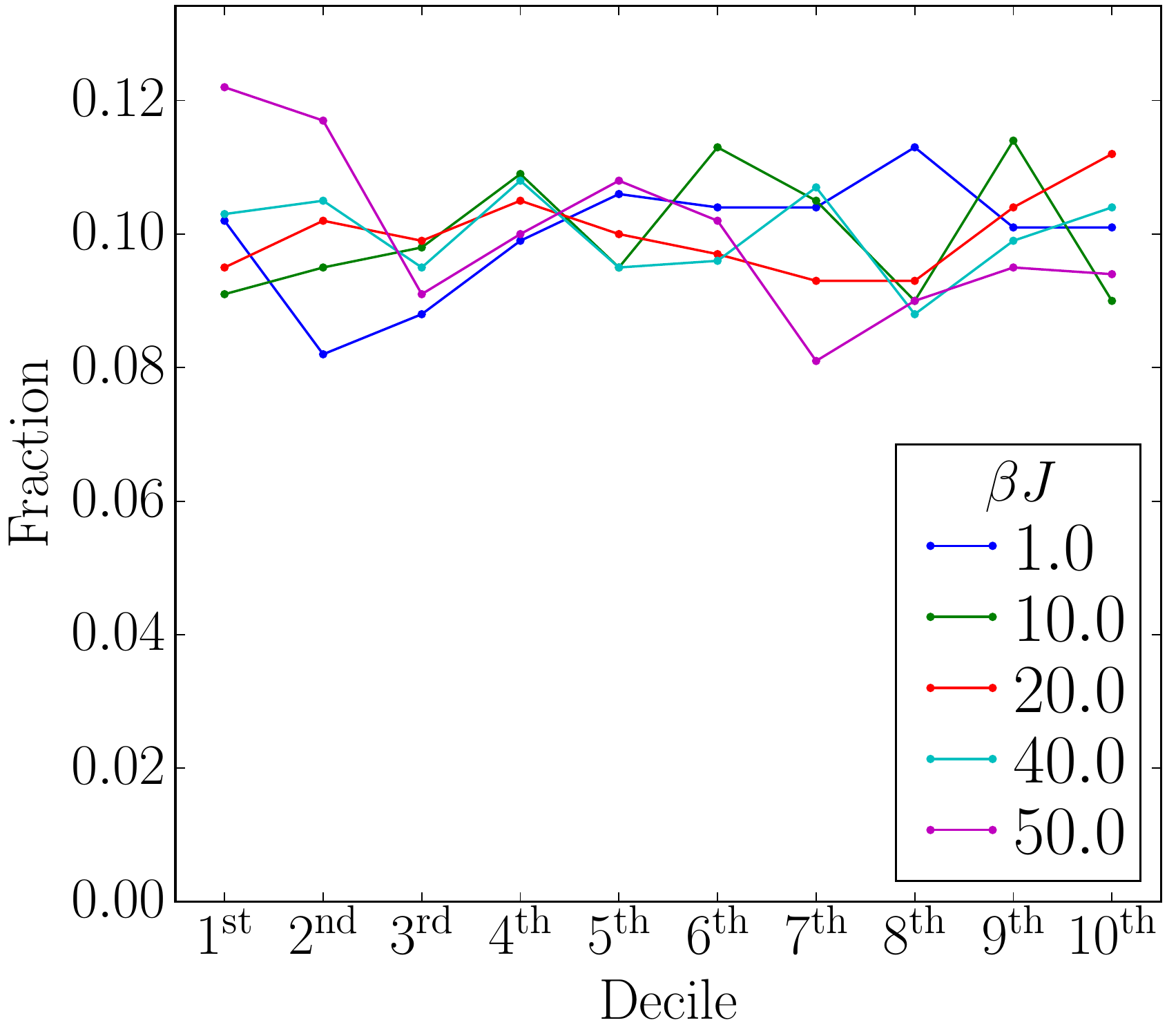}
\caption{\label{fig:calibration}{\em Accuracy and precision; Bayesian}. The fraction, over 1000 simulated experiments, of experiments in which the true value of $\beta$ lies in the deciles of $\PP(\beta)$ predicted using Bayesian analysis. The flatness of the plots, especially for the lower temperatures, reveals that our null priors are well calibrated. All parameters are identical to those for \fir{fig:lowtempfreq}.}
\end{figure}


\section{Theoretical values for $\chi_Q(u)$}
\label{sec:chicalc}
In this part of the supplemental material, we provide details of the two methods used to calculate the characteristic function (as well as other relevant quantities) of our example system. Specifically, we calculate
\begin{equation}
\label{eq:CharacteristicFunction}
\chi_Q (u) = \tr \left\{ \hat{U}^\dagger_Q \hat{U}^\dagger ( \lambda_Q(\tau) , u) \hat{U}_Q \hat{U} ( \lambda_Q(0) , u) \hat{\rho}_\beta (\lambda_Q(0)) \right\} , 
\end{equation}
where 
\begin{align}
\hat{U}_Q = & \TT \exp \left [ - \ii \int_0^{\tau} \dds t \hat{H} (\lambda_Q(t)) \right] , \label{eq:UnitaryEvolution} \\
\hat{U} ( \lambda , u) = & \ee^{-\ii \hat{H}(\lambda) u} ,  \nonumber \\
\hat{\rho}_\beta (\lambda) = & \ee^{-\beta \hat{H} (\lambda)} / \ZZ_\beta (\lambda) ,  \nonumber \\
\ZZ_\beta (\lambda) = & \tr \{ \exp [ - \beta \hat{H} (\lambda) ]\} ,  \nonumber \\
\hat{H} (\lambda) = & \hat{H}_S +  \lambda \hat{V} . \nonumber 
\end{align}
We begin by discussing the Bogoliubov treatment, relevant to the superfluid phase, and then move on to the tensor network approach that is necessary when interactions are stronger.

\subsection{Bogoliubov treatment}
\subsubsection{The free energy}
In the Bogoliubov treatment~\cite{VanOosten2001} we have the simplified Hamtilonian $\hat{H} (\lambda) = \hat{H}_S +  \lambda \hat{V} $, where
\begin{align}
\an{H}_S = & \sum_k \omega_k \cre{b}_k \an{b}_k ,  \nonumber  \\
\an{V} = & \eta n + \sum_k ( \eta^\ast_k \cre{b}_k + \eta_k \an{b}_k ) .  \nonumber
\end{align}
Here $\cre{b}_k$ and $\an{b}_k$, introduced in the main text along with the other terms, satisfy bosonic commutation relations. 

This Hamiltonian can be diagonalized by defining a new displaced creation operator $\cre{d}_k = \cre{b}_k + \lambda \eta_k/\omega_k$, and its annihilating conjugate $\an{d}_k$, leading to 
\begin{align}
\hat{H} (\lambda) = \sum_k \omega_k \cre{d}_k \an{d}_k + \lambda \eta \left( n - \lambda \sum_k \frac{|\eta_k|^2}{\eta \omega_k} \right ) . \nonumber 
\end{align}
We can thus immediately write the free energy as
\begin{align}
F (\lambda) &= -\frac{1}{\beta} \ln \ZZ_\beta (\lambda)  \nonumber \\
&= \lambda \eta \left( n - \lambda \sum_k \frac{|\eta_k|^2}{\eta \omega_k} \right ) + F(0) ,  \nonumber \\
F(0) &= -\frac{1}{\beta} \ln \tr \left \{ \exp \left[ - \beta \sum_k \omega_k \cre{d}_k \an{d}_k \right] \right \}  \nonumber  \\
&= \frac{1}{\beta} \sum_k \ln ( 1 - \ee^{-\beta \omega_k} ) .  \nonumber 
\end{align}

\subsubsection{The characteristic function}
Our first step in evaluating $\chi_Q (u)$ [\eqr{eq:CharacteristicFunction}] is to simplify by absorbing any displacement of the Bogoliubov phonons present in the initial state $\hat{\rho}_\beta (\lambda_Q(0))$ into the Hamiltonian $\hat{H} (\lambda)$. We do this using a method similar to that used for the free energy. We define a new displaced creation operator $\cre{d}_k = \cre{b}_k + \lambda_Q (0) \eta_k/\omega_k$ and its annihilating conjugate $\an{d}_k$. In terms of these operators, we re-write the Hamiltonian as
\begin{align}
\hat{H} (\lambda) = & \hat{H}' (\lambda') = \hat{H}'_S +  \lambda' \hat{V}' ,  \nonumber \\
\lambda' = & \lambda - \lambda_Q(0) ,  \nonumber \\
\an{H}'_S = & \sum_k \omega_k \cre{d}_k \an{d}_k ,  \nonumber \\
\an{V}' = & \eta n' + \sum_k ( \eta^\ast_k \cre{d}_k + \mathrm{h.c.} ) ,  \nonumber \\
n' = & n - 2 \lambda_Q (0) \sum_k \frac{|\eta_k|^2}{\eta \omega_k} . \nonumber 
\end{align}
Here $n'$ represents the reduced background density due to the initial perturbation, and we have omitted a constant term $\lambda_Q (0) \eta ( n - \lambda_Q (0) \sum_k |\eta_k|^2 /\eta \omega_k )$ representing its energy.

We now have
\begin{equation}
\label{eq:CharacteristicFunction2}
\chi_Q (u) = \tr \left\{ \hat{U}^{\prime \dagger}_Q \hat{U}^{\prime \dagger} ( \lambda'_Q (\tau) , u) \hat{U}'_Q \hat{\rho}'_\beta (0) \right\} , \nonumber 
\end{equation}
where 
\begin{align}
\hat{U}'_Q = & \TT \exp [ - \ii \int_0^{\tau} \dds t \hat{H}' (\lambda'_Q(t)) ] ,  \nonumber \\
\hat{U}' ( \lambda' , u) = & \ee^{-\ii \hat{H}'(\lambda') u} ,  \nonumber \\
\hat{\rho}'_\beta (\lambda') = & \ee^{-\beta \hat{H}' (\lambda')} / \ZZ'_\beta (\lambda') ,  \nonumber \\
\ZZ'_\beta (\lambda') = & \tr \{ \exp [ - \beta \hat{H}' (\lambda') ]\}.  \nonumber 
\end{align}
Note that, by design, $\lambda'_Q(0) = 0$ and thus we have used $\hat{U}' ( \lambda'_Q(0) , u) = \hat{U}' ( 0 , u) = 1$. In what follows, for clarity, we drop the primes. 

Our second step is to move to the interaction picture, whereupon
\begin{equation}
\label{eq:CharacteristicFunction3}
\chi_Q (u) = \tr \left\{ \tilde{U}^{ \dagger}_Q(0) \tilde{U}^{\dagger} ( \lambda_Q(\tau) , u, \tau) \tilde{U}_Q (u) \an{\rho}_\beta (0) \right\} .  \nonumber 
\end{equation}
Here the tilde indicates an operator in the interaction picture, specifically
\begin{align}
\tilde{U}_Q(t) =& \TT \exp \left [ - \ii \int_0^{\tau} \dds t' \lambda_Q (t') \tilde{V} (t'+t) \right ] ,  \nonumber \\ 
\tilde{U} (\lambda, u, t) =& \TT \exp \left [ - \ii \lambda \int_{0}^{u} \dds t' \tilde{V} (t' + t) \right ] ,  \nonumber \\ 
\tilde{V} (t) =& \ee^{\ii \an{H}_S t} \hat{V} \ee^{-\ii \an{H}_S t}  \nonumber \\
=& \eta n + \sum_k ( \eta^\ast_k \cre{d}_k \ee^{\ii \omega_k t} + \mathrm{h.c.} ) .  \nonumber 
\end{align}

The third step is to simplify the time-ordered exponentials. For this we appeal to the Magnus expansion
\begin{align}
\tilde{U} =& \TT \exp \left [ - \ii \int_0^{t} \dds t' \lambda (t') \tilde{V} (t') \right ] = \ee^{-\ii \tilde{A}} ,  \nonumber 
\end{align}
in terms of Hermitian operators
\begin{align}
\tilde{A} =& \tilde{A}^{[1]} + A^{[2]} + \cdots ,  \nonumber \\
\tilde{A}^{[1]} =& \int_0^{t} \dds t' \lambda (t') \tilde{V} (t') ,  \nonumber \\
A^{[2]} =& \frac{-\ii}{2} \int_0^{t} \dds t' \int_0^{t'} \dds t'' \lambda (t') \lambda (t'') [ \tilde{V} (t'), \tilde{V} (t'')] ,  \nonumber \\
\vdots \nonumber 
\end{align}
The simplifying feature of the Bogoliubov Hamiltonian, which we will use repeatedly, is that the commutator
\begin{equation}
[ \tilde{V} (t), \tilde{V} (t')] = -2 \ii \sum_k |\eta_k|^2 \sin (\omega_k (t-t')) ,  \nonumber 
\end{equation}
is a pure imaginary $c$-number. Hence we have omitted the tilde from $A^{[2]}$ to highlight that it is a real c-number only. This also ensures that all terms $\tilde{A}^{[m]}$ for $m>2$ vanish from the Magnus expansion.  The result is that
\begin{align}
\chi_Q (u) = & \ee^{\ii A^{[2]} (\lambda_Q (\tau),u)}  \nonumber \\ 
 & \times \tr \left\{ \ee^{\ii \tilde{A}^{[1]}_Q (0)} \ee^{\ii \tilde{A}^{[1]} ( \lambda_Q (\tau) , u, \tau)}  \ee^{-\ii \tilde{A}^{[1]}_Q (u)} \an{\rho}_\beta (0) \right\} .  \nonumber 
\end{align}
Here, we have used a simplification of the form $\ee^{\ii A^{[2]}_Q (0)} \ee^{-\ii A^{[2]}_Q (u)} = 1$, and the other integrals appearing in the expression are as follows
\begin{align}
\tilde{A}^{[1]}_Q(t) =& \int_0^{\tau} \dds t' \lambda_Q (t') \tilde{V} (t'+t)   \nonumber \\
=& \eta n \int_0^{\tau} \dds t' \lambda_Q (t')  \nonumber \\ 
&+ \sum_k \left( \frac{\eta_k^\ast}{\omega_k} \Lambda_{Q k}^\ast \cre{d}_k \ee^{\ii \omega_k t} + \mathrm{h.c.} \right) , \nonumber  \\
\Lambda_{Q k} =& \omega_k \int_0^{\tau} \dds t \lambda_Q (t) \ee^{-\ii \omega_k t} ,  \nonumber \\
\tilde{A}^{[1]} ( \lambda , u, t) =& \lambda \int_0^{u} \dds t' \tilde{V} (t' + t)  \nonumber \\
=& \eta n \lambda u  \nonumber \\ 
&+ \sum_k \left( \frac{\eta_k^\ast}{\omega_k} \Lambda^\ast_k (\lambda,u) \cre{d}_k \ee^{\ii \omega_k t} + \mathrm{h.c.} \right) ,  \nonumber \\
\Lambda_k (\lambda,u) =& \omega_k \lambda \int_0^{u} \dds t \ee^{-\ii \omega_k t}  \nonumber \\
=& -\ii \lambda (1-\ee^{-\ii \omega_k u}) ,  \nonumber \\
A^{[2]} ( \lambda , u) =& - \lambda^2 \sum_k \left |\frac{\eta_k}{\omega_k} \right|^2 \left ( \omega_k u - \sin (\omega_k u) \right ) . \nonumber 
\end{align}

Having simplified each time-ordered exponential, our third step is to combine them using the Baker-Campbell-Hausdorff formula
\begin{align}
\ee^{\tilde{A}} \ee^{\tilde{B}} = \ee^{\tilde{A}+ \tilde{B} + [\tilde{A}, \tilde{B}]/2},  \nonumber 
\end{align}
for the case that $[\tilde{A}, \tilde{B}]$ is a $c$-number. Explicitly, we use
\begin{align}
& \ee^{\ii \tilde{A}^{[1]}_Q(0)} \ee^{\ii \tilde{A}^{[1]} ( \lambda_Q (\tau) , u, \tau)}  \ee^{-\ii \tilde{A}^{[1]}_Q(u)}  \nonumber \\ 
& = \ee^{\ii (\tilde{A}^{[1]}_Q(0)-\tilde{A}^{[1]}_Q (u)) +\ii \tilde{A}^{[1]} ( \lambda_Q(\tau) , u, \tau)}   \nonumber \\
& \quad \times \exp \Big \{ -\frac{1}{2} \Big ( [\tilde{A}^{[1]}_Q(u),\tilde{A}^{[1]}_Q (0)]  \nonumber 
\\ & \qquad \qquad \qquad \quad  +[\tilde{A}^{[1]}_Q(u),\tilde{A}^{[1]} ( \lambda_Q(\tau) , u ,\tau)]  \nonumber 
\\ & \qquad \qquad \qquad \quad  +[\tilde{A}^{[1]}_Q(0),\tilde{A}^{[1]} ( \lambda_Q(\tau) , u, \tau)] \Big ) \Big \}  , \nonumber 
\end{align}
together with
\begin{align}
&[\tilde{A}^{[1]}_Q(t),\tilde{A}^{[1]}_Q (t')]  \nonumber \\
&=  -2\ii \sum_k \left |\frac{\eta_k}{\omega_k} \right|^2 | \Lambda_{Q k} |^2 \sin ( \omega_k (t-t')) ,  \nonumber \\
&[\tilde{A}^{[1]}_Q(t),\tilde{A}^{[1]} ( \lambda , u, t') ]  \nonumber \\
&=  -2\ii \sum_k \left |\frac{\eta_k}{\omega_k} \right|^2 \Im \left \{ \Lambda^\ast_{Q k} \Lambda_k (\lambda,u) \ee^{\ii \omega_k (t-t')} \right \} . \nonumber 
\end{align}

This leaves us with
\begin{equation}
\label{eq:chisemifinal}
\begin{aligned}
\chi_Q (u) = & \exp \left( \ii \eta n \lambda_Q (\tau) u  \right) \\
& \times \exp \left( -\ii \lambda_Q^2(\tau) \sum_k \left |\frac{\eta_k}{\omega_k} \right|^2 \left ( \omega_k u - \sin (\omega_k u) \right ) \right) \\
& \times \exp \left( - \ii \sum_k \left |\frac{\eta_k}{\omega_k} \right|^2 h_{Q k} (u) \right) \\
& \times \tr \left \{ \exp \left( - \ii \sum_k \left ( \frac{\eta^\ast_k}{\omega_k} g^\ast_{Q k} (u) \cre{d}_k + \mathrm{h.c.} \right) \right) \an{\rho}_\beta (0) \right \} ,
\end{aligned}
\end{equation}
where
\begin{align}
h_{Q k} (u) = & - | \Lambda_{Q k} |^2 \sin ( \omega_k u)  \nonumber \\ 
 & - \Im \Big \{ \Lambda^\ast_{Q k} \left (  \ee^{\ii \omega_k u} + 1 \right) \Lambda_k (\lambda_Q(\tau),u) \ee^{-\ii \omega_k \tau} \Big \}  \nonumber \\
 = & H_{Qk} \sin ( \omega_k u) ,  \nonumber \\
 H_{Qk} = & -| \Lambda_{Q k} |^2 -2 \lambda_Q(\tau) \Im \Big \{ \Lambda^\ast_{Q k} \ee^{-\ii \omega_k \tau} \Big \} ,  \nonumber \\
g_{Q k} (u) = &  -\Lambda_{Q k} \left( 1 - \ee^{-\ii \omega_k u} \right) - \Lambda_k (\lambda_Q(\tau),u) \ee^{-\ii \omega_k \tau}  \nonumber \\
= & G_{Q k} \left( 1 - \ee^{-\ii \omega_k u} \right) , \nonumber \\
G_{Q k} = &  - \Lambda_{Q k} + \ii \lambda_Q(\tau) \ee^{-\ii \omega_k \tau} . \nonumber 
\end{align}
The final step is to evaluate the trace $\tr \{ \cdot \hat{\rho}_\beta (0) \}$, or expected value $\langle \cdot \rangle$ with respect to state $\hat{\rho}_\beta (0)$. We use that the exponent of the first term in the trace contains only terms that are linear in $\cre{d}_k$ and $\an{d}_k$. The state $\hat{\rho}_\beta (0)$ with respect to which the expected value is taken, the other term in the trace, consists of a mixture of different occupations of these phonon modes, with all non-number conserving combinations of  $\cre{d}_k$ and $\an{d}_k$ thus having zero expected value and $\langle (\cre{d}_k \an{d}_k)^m \rangle = (n_k)^m$ with $n_k = ( \exp(\beta \omega_k) - 1)^{-1}$ the mean occupation. This means that for an operator formed for linear combinations of $\cre{d}_k$ and $\an{d}_k$, we have
\begin{align}
& \left \langle \exp \left( - \ii \sum_k \left ( \frac{\eta^\ast_k}{\omega_k} g^\ast_{Q k} (u) \cre{d}_k + \mathrm{h.c.} \right) \right) \right \rangle  \nonumber \\ 
=& \exp \left( - \frac{1}{2} \left \langle \left ( \sum_k \left ( \frac{\eta^\ast_k}{\omega_k} g^\ast_{Q k} (u) \cre{d}_k + \mathrm{h.c.} \right) \right)^2 \right \rangle \right)  , \nonumber 
\end{align}
and the particular expectation value is
\begin{align}
& \left \langle \left ( \sum_k \left ( \frac{\eta^\ast_k}{\omega_k} g^\ast_{Q k} (u) \cre{d}_k + \mathrm{h.c.} \right) \right)^2 \right \rangle  \nonumber \\
=&  \sum_k \left |\frac{\eta_k}{\omega_k} \right|^2 \left | g_{Q k} (u) \right|^2 \coth \left( \frac{ \beta \omega_k}{2} \right) , \nonumber 
\end{align}
where we note that
\begin{align}
\left | g_{Q k} (u) \right |^2 = & 4 \left | G_{Q k} \right |^2 \sin^2 \left ( \frac{\omega_k u}{2} \right) . \nonumber 
\end{align}

Inserting this back into \eqr{eq:chisemifinal} we arrive at our final expression
\begin{widetext}
\begin{align}
i \ln \chi_Q (u) =& - \eta n \lambda_Q (\tau) u + \sum_k \left |\frac{\eta_k}{\omega_k} \right|^2 \left( \lambda^2_Q (\tau) \omega_k u +  (H_{Qk} - \lambda^2_Q (\tau)) \sin ( \omega_k u) \right) - 2 \ii \sum_k \left |\frac{\eta_k}{\omega_k} \right|^2 \left | G_{Q k} \right |^2 \frac{\sin^2 \left ( \frac{\omega_k u}{2} \right) }{\tanh \left( \frac{ \beta \omega_k}{2} \right)} .  \nonumber 
\end{align}
\end{widetext}

\subsubsection{The work distribution}
As discussed above, the cumulants $\kappa_m$ of the work distribution $P_Q(W)$ are related to the derivatives of the logarithm of its characteristic function, by \eqr{eq:cumulants}.
We may use this to calculate all cumulants of the work distribution, but here we report only the first three
\begin{align}
\mu_Q = \kappa_{Q1} = & \lambda_Q (\tau) \eta n - \sum_k \frac{\left | \eta_k \right|^2}{\omega_k} H_{Qk} , \nonumber \\
\sigma_Q^2 = \kappa_{Q2} = & \sum_k \left | \eta_k \right|^2 \left | G_{Qk} \right|^2 \coth \left( \frac{ \beta \omega_k}{2} \right) , \nonumber \\
\kappa_{Q3} = & - \sum_k \left | \eta_k \right|^2 \omega_k H_{Qk} .\nonumber 
\end{align}

\subsubsection{Symmetry}

A close inspection of the results reveals that reversing the quench has no effect on either $H_{Qk}$ or $\left | G_{Q k} \right |^2$ and thus also on the cumulants $\kappa_{Qm}$ for $m>1$. This represents the fact that in the Bogoliubov description the deviations of the condensate, assumed small, do not to interact resulting in a symmetry between repulsive and attractive interactions. This means that the only difference between the two characteristic functions lies in the phase $\eta n \lambda_Q (\tau) u$. The effect of this is merely to shift the associated probability distributions by the amount $\eta n \lambda_Q (\tau)$, as we can see in our expression for $\mu_Q$. Evaluating these phase shifts, we find that, for the Bogoliubov case, the forward and backward distributions are identical up to a relative shift $\mu_F - \mu_B = 2\Delta F$, where $\Delta F = F(\lambda_f) - F(\lambda_i)$ is the free energy difference.

This symmetry could be exploited to provide a better estimate for $\beta$. However, we do not do this here as we wish to keep our estimation protocol general, so that it is applicable to situations where this symmetry does not exist or is not known to exist.

\subsection{Tensor network theory}

\begin{figure}
\includegraphics[scale=0.2]{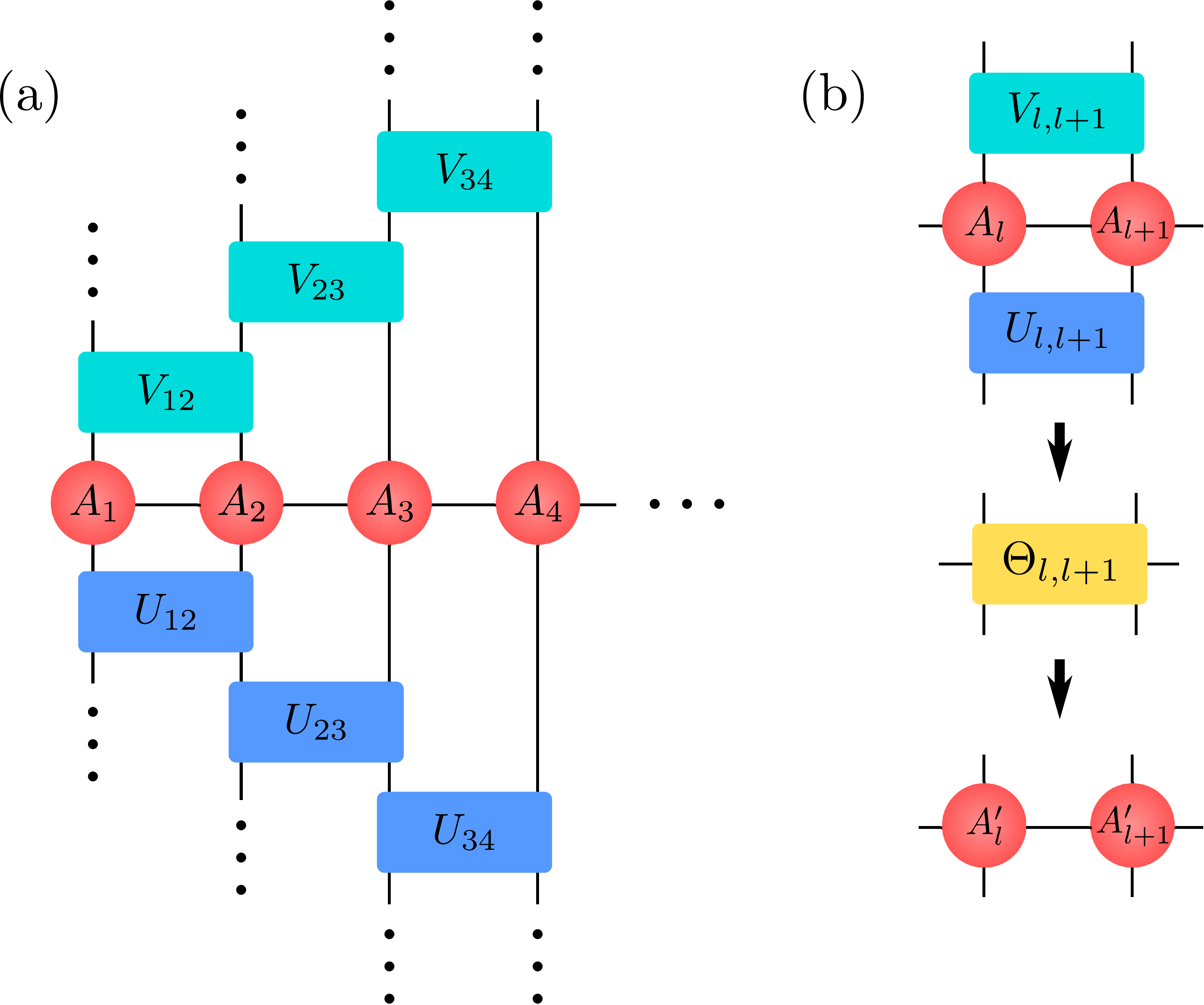}
\caption{(a) Schematic depicting one part of a tensor network representing the time evolution of an MPO (red circles) generated by a product of two-site gates $U_{l,l+1}$ and $V_{l,l+1}$ acting on the MPO from the left and right, respectively. (b) Time evolution proceeds by contraction of nearest-neighbor pairs of MPO tensors $A_lA_{l+1}$ with two-site gates above and below, resulting in a new tensor $\Theta_{l,l+1}$. This tensor is then decomposed into a product of tensors $A'_l A'_{l+1}$ via an SVD. Retaining only the largest $D$ singular values in the SVD maintains an efficient MPO representation at each time step. \label{tensorNetworkAppendix}}
\end{figure}

In regimes where the condition necessary for the validity of Bogoliubov theory $U n\ll J$ is not satisfied, we numerically calculate the characteristic function using a tensor network representation. We proceed by writing the quantum state as a matrix product operator (MPO) \cite{Verstraete2004prl,Zwolak2004prl} and then performing imaginary- and real-time evolution using the time-evolving block decimation (TEBD) algorithm \cite{Vidal2004prl,White2004prl}. In the following, we concisely outline our method. See Ref.~\cite{Schollwock2011aop} for a detailed pedagogical introduction to tensor-network algorithms.

The state of a quantum lattice system with $M$ sites and local Hilbert space dimension $d$ can be represented in MPO form
\begin{equation}
\label{mpoRepresentation}
\begin{aligned}
\hat{\rho} = \sum_{i_1,\ldots,i_M = 1}^d\;\sum_{j_1,\ldots,j_M = 1}^d \mathrm{tr} &\left [ A^{(i_1j_1)}_1  \cdots A^{(i_Lj_M)}_M\right ]  \\ & \times \lvert i_1\ldots i_M\rangle\langle j_1\ldots j_M\rvert.
\end{aligned}
\end{equation}
Here, the matrix elements of the density operator are given by the trace over a product of matrices $A^{(i_l j_l)}_l$ of maximum dimension $D\times D$, where the bond dimension $D$ quantifies the correlations between lattice sites. The states $\{ \lvert i_l\rangle\}$ constitute a complete orthonormal basis for the Hilbert space on lattice site $l$. The set of matrices $A^{(i_l j_l)}_l$ may also be considered as elements of a single combined tensor $A_l$ of dimension $D \times d \times D \times d$. Together the tensors $A_l$ can represent an arbitrary quantum state, for sufficiently large $d$ and $D$. In a bosonic system, however, both the physical dimension $d$ and the bond dimension $D$ must be truncated, in general, in order to obtain a tractable numerical representation. In our calculations we use $d = 4$ and $D=200$. 

The time-evolution operator over a small time step $\delta t$ is approximated by a product of two-site gates using a second-order Suzuki-Trotter ``staircase" decomposition, see Ref.~\cite{Johnson2010pre} for details. The tensor network representing the time evolution of the quantum state is depicted schematically in Fig.~\ref{tensorNetworkAppendix}(a). The TEBD algorithm proceeds by sweeping across this tensor network and applying two-site gates sequentially to pairs of nearest-neighbor tensors $A_l A_{l+1}$ appearing in the MPO representation of the quantum state in \eqr{mpoRepresentation}. The result is a new tensor $\Theta_{l,l+1}$ formed by contracting the $A$ tensors with two-site gates above and below. A singular value decomposition (SVD) is then performed on $\Theta_{l,l+1}$, resulting in a new pair of tensors $A'_l A'_{l+1}$ [Fig.~\ref{tensorNetworkAppendix}(b)]. Only the largest $D$ singular values at most are retained after the SVD, so that an efficient MPO representation of the quantum state is maintained at each time step.  Repeating this procedure across the entire system over $N$ small time steps $\delta t$ leads to the desired numerical evolution over a time duration $t = N\delta t$.

In order to find the initial thermal state, we use the identity
\begin{equation}
\label{imaginaryTimeEvolution}
\hat{\rho}_\beta(\lambda) = \frac{1}{\mathcal{Z}_\beta(\lambda)} \mathrm{e}^{-\beta \hat{H} /2} \hat{\mathds{1}} \mathrm{e}^{-\beta \hat{H}/2}. \nonumber
\end{equation}
The right-hand side of this equation can be calculated using the TEBD algorithm as outlined above, after a Wick rotation to imaginary time $t\to-\mathrm{i}\beta/2$ and taking the initial state to be the system-wide identity operator $\hat{\mathds{1}}$. The MPO representation of the identity operator is given simply by $A_{l}^{(i_lj_l)} = \delta_{i_lj_l}$.

The characteristic function [\eqr{eq:CharacteristicFunction}] may then be calculated by performing real-time evolution to find the operator
\begin{equation}
\label{correlationFunction}
\hat{X}_Q(u) = \hat{U}_Q \mathrm{e}^{-\mathrm{i}u\hat{H}(\lambda_Q(0))} \hat{\rho}_\beta(\lambda_Q(0))  \hat{U}_Q^\dagger \mathrm{e}^{\mathrm{i}u\hat{H}(\lambda_Q(\tau))}, \nonumber
\end{equation}
such that $\chi_Q(u) = \mathrm{tr}\left \{ \hat{X}_Q(u)\right \}$. The unitary $\an{U}_Q =\TT \exp [ - \ii \int_0^{\tau} \dd t \hat{H} (\lambda_Q(t) ) ]$  [\eqr{eq:UnitaryEvolution}] is performed by discretizing the quench path into steps $\lambda_{Qm} = \lambda_Q(m \delta t)$, with $m = 1,\ldots,M$ and $\tau = M\delta t$. Discrete time evolution under the TEBD algorithm reproduces the quench unitary, since
\begin{equation}
\label{Uqdiscrete}
\an{U}_Q \approx \prod_{m=M}^1 \mathrm{e}^{-\mathrm{i} \delta t H(\lambda_{Qm})} , \nonumber
\end{equation}
for sufficiently small $\delta t$. 

In order to ensure numerical stability, each matrix $A_l^{(i_lj_l)}$ is divided by its matrix norm (the square root of the sum of its elements) after each time step. The accumulated product of these norms then multiplies expectation values to give the correct final result.

\end{document}